\documentclass{einformatyka}

\usepackage[utf8]{inputenc}
\usepackage{graphicx}
\usepackage{color,soul}

\begin{document}

\title{Software Startups - A Research Agenda}

\begin{articleinfo}
	
\begin{authorgroup}
	\begin{author}
		\firstname{Michael}
		\surname{Unterkalmsteiner}
		\orgname{Blekinge Institute of Technology, Sweden}
		\email{mun@bth.se}
	\end{author}	
	\begin{author}
		\firstname{Pekka}
		\surname{Abrahamsson}
		\orgname{Norwegian University of Science and Technology, Norway}
		\email{pekkaa@ntnu.no}
	\end{author}
	\begin{author}
		\firstname{XiaoFeng}
		\surname{Wang}
		\orgname{Free University of Bolzano-Bozen, Italy}
		\email{xiaofeng.wang@unibz.it}
	\end{author}	
	\begin{author}
		\firstname{Anh}
		\surname{Nguyen-Duc}
		\orgname{Norwegian University of Science and Technology, Norway}
		\email{anhn@idi.ntnu.no}
	\end{author}
	\begin{author}
		\firstname{Syed}
		\surname{Shah}
		\orgname{SICS, Sweden}
		\email{shah@sics.se}
	\end{author}	
	\begin{author}
		\firstname{Sohaib Shahid}
		\surname{Bajwa}
		\orgname{Free University of Bolzano-Bozen, Italy}
		\email{bajwa@inf.unibz.it}
	\end{author}
	\begin{author}
		\firstname{Guido H.}
		\surname{Baltes}
		\orgname{Lake Constance University, Germany}
		\email{guido.baltes@cetim.org}
	\end{author}
	\begin{author}
		\firstname{Kieran}
		\surname{Conboy}
		\orgname{National University of Ireland Galway, Ireland}
		\email{kieran.conboy@nuigalway.ie}
	\end{author}
	\begin{author}
		\firstname{Eoin}
		\surname{Cullina}
		\orgname{National University of Ireland Galway, Ireland}
		\email{eoin.cullina@outlook.com}
	\end{author}
	\begin{author}
		\firstname{Denis}
		\surname{Dennehy}
		\orgname{National University of Ireland Galway, Ireland}
		\email{denis.dennehy@nuigalway.ie}
	\end{author}
	\begin{author}
		\firstname{Henry}
		\surname{Edison}
		\orgname{Free University of Bolzano-Bozen, Italy}
		\email{henry.edison@inf.unibz.it}
	\end{author}
	\begin{author}
		\firstname{Carlos}
		\surname{Fernandez-Sanchez}
		\orgname{Universidad Politécnica de Madrid, Spain}
		\email{carlos.fernandez@upm.es}	
	\end{author}
	\begin{author}
		\firstname{Juan}
		\surname{Garbajosa}
		\orgname{Technical University of Madrid, Spain}
		\email{jgs@eui.upm.es}
	\end{author}
	\begin{author}
		\firstname{Tony}
		\surname{Gorschek}
		\orgname{Blekinge Institute of Technology, Sweden}
		\email{tgo@bth.se}
	\end{author}
	\begin{author}
		\firstname{Eriks}
		\surname{Klotins}
		\orgname{Blekinge Institute of Technology, Sweden}
		\email{ekx@bth.se}
	\end{author}
	\begin{author}
		\firstname{Laura}
		\surname{Hokkanen}
		\orgname{Tampere University of Technology, Finland}
		\email{laura.hokkanen@tut.fi}
	\end{author}
	\begin{author}
		\firstname{Fabio}
		\surname{Kon}
		\orgname{University of São Paulo, Brazil}
		\email{fabio.kon@ime.usp.br}
	\end{author}	
	\begin{author}
		\firstname{Ilaria}
		\surname{Lunesu}
		\orgname{University of Cagliari, Italy}
		\email{ilaria.lunesu@diee.unica.it}
	\end{author}
	\begin{author}
		\firstname{Michele}
		\surname{Marchesi}
		\orgname{University of Cagliari, Italy}
		\email{michele@diee.unica.it}
	\end{author}
	\begin{author}
		\firstname{Lorraine}
		\surname{Morgan}
		\orgname{National University of Ireland Maynooth, Ireland}
		\email{lorraine.morgan@nuim.ie}
	\end{author}
	\begin{author}
		\firstname{Markku}
		\surname{Oivo}
		\orgname{University of Oulu, Finland}
		\email{markku.oivo@oulu.fi}
	\end{author}
	\begin{author}
		\firstname{Christoph}
		\surname{Selig}
		\orgname{Hochschule Konstanz, Germany}
		\email{cselig@htwg-konstanz.de}
	\end{author}
	\begin{author}
		\firstname{Pertti}
		\surname{Seppänen}
		\orgname{University of Oulu, Finland}
		\email{pertti.seppanen@oulu.fi}
	\end{author}
	\begin{author}
		\firstname{Roger}
		\surname{Sweetman}
		\orgname{National University of Ireland Galway, Ireland}
		\email{roger.sweetman@nuigalway.ie}
	\end{author}
	\begin{author}
		\firstname{Pasi}
		\surname{Tyrväinen}
		\orgname{University of Jyväskylä, Finland}
		\email{pasi.tyrvainen@jyu.fi}	
	\end{author}	
	\begin{author}
		\firstname{Christina}
		\surname{Ungerer}
		\orgname{Hochschule Konstanz, Germany}
		\email{christina.ungerer@htwg-konstanz.de}
	\end{author}
	\begin{author}
		\firstname{Agustin}
		\surname{Yagüe}
		\orgname{Universidad Politécnica de Madrid, Spain}
		\email{ayague@etsisi.upm.es}	
	\end{author}
\end{authorgroup}

\begin{abstract}[en]
Software startup
companies develop innovative, software-intensive products within limited time frames and 
with few resources, searching for sustainable and scalable 
business models. Software startups are quite distinct from traditional 
mature software companies, but also from micro-, small-, and medium-sized 
enterprises, introducing new challenges relevant for software engineering 
research. This paper's research agenda focuses on software engineering in 
startups, identifying, in particular, 70+ research questions 
in the areas of supporting startup engineering activities, startup evolution 
models and patterns, ecosystems and innovation hubs, human aspects in software 
startups, applying startup concepts in non-startup environments, and 
methodologies and theories for startup research. We connect and motivate this 
research agenda with past studies in software startup research, while pointing out 
possible future directions. While all authors of this research agenda have their main 
background in Software Engineering or Computer Science, their interest in software 
startups broadens the perspective to the challenges, but also to the opportunities that 
emerge from multi-disciplinary research. Our audience is therefore primarily software 
engineering researchers, even though we aim at stimulating collaborations and research 
that crosses disciplinary boundaries. We believe that with this research agenda we cover 
a wide spectrum of the software startup industry current needs.
\end{abstract}

\end{articleinfo}

\section{Introduction}
Researchers are naturally drawn to complex phenomena that challenge their 
understanding of the world. Software startup companies are an intriguing phenomenon, 
because they develop innovative software-intensive\footnote{ISO 
42010:2011~\cite{_iso/iec/ieee_2011} defines 
software-intensive systems as ``any system where software contributes essential 
influences to the design, construction, deployment, and evolution of the system as a 
whole'' to encompass ``individual applications, systems in the traditional sense, 
subsystems, systems of systems, product lines, product families, whole enterprises, and 
other aggregations of interest''.} products under 
time constraints and with a lack of resources~\cite{paternoster_software_2014}, 
and constantly search for sustainable and scalable business models. Over the past few 
years, software startups have garnered increased research interest in the Software 
Engineering (SE) community.
 
While one could argue that software startups represent an exceptional 
case of how software products are developed and brought to the market, several factors 
suggest a broader impact. From an economical perspective, startups contribute 
considerably to overall wealth and progress by creating jobs and 
innovation~\cite{sba2014}. Digital software startups\footnote{In our article, digital 
startups refer specifically to startups in which the business value of the solution is 
created by means of software~\cite{nambisan_digital_2016}.} are responsible for 
an astonishing variety of services and products~\cite{Economist20140118}. In the farming 
sector, venture investment in so-called ``AgTech'' start-ups reached \$2.06 billion in 
just the first half of 2015; this figure neared the \$2.36 billion raised during the 
whole of 2014~\cite{wmf2015}. From an innovation perspective, startups often pave the way 
for the introduction of even more new and disruptive 
innovations~\cite{srinivasan_venture_2014}. Kickstarter is 
changing the retail and finance industries, Spotify is offering a new way to 
listen to and purchase music, and Airbnb is reinventing the hospitality 
industry~\cite{Shontell2012}. From an engineering perspective, startups must inventively 
apply existing knowledge in order to open up unexpected avenues 
for improvement~\cite{giardino_software_2016}; e.g., they must provide education for full 
stack engineers, develop techniques for continuous lightweight requirements 
engineering, or develop strategies to control technical debt.

Despite these promising conditions, software startups face challenges to survival, even 
in contexts where they play a key role in developing new technology and markets, such as 
cloud computing~\cite{Economist20150919}. These challenges may arise because, while 
developing a product can be easy, selling it can be quite 
difficult~\cite{Economist20140118-2}. Software startups face other challenges, such as 
developing cutting-edge products, acquiring 
paying customers, and building entrepreneurial teams~\cite{giardino_key_2015}. 
Such diverse factors underscore the need to conduct research on software 
startups, which will benefit both scholarly communities and startup leaders.

This paper's research agenda is driven by past and current work on software startups. We 
outline the various research tracks to provide a snapshot of ongoing work and to preview 
future research, creating a platform for identifying collaborations with both research 
and startup environments and ecosystems. This effort is not a one-way path. 
We have therefore founded a research network, the Software Startup Research Network 
(SSRN)\footnote{https://softwarestartups.org}, which enables 
interactions and collaborations among researchers and interested startups. 
SSRN envisions to: (1) spread novel research findings in the context of 
software startups; and (2) inform entrepreneurs with necessary knowledge, 
tools and methods that minimize threats and maximize opportunities for success.
As part of the network initiatives, an International Workshop of Software 
Startups was established in 2015. The first edition of the workshop was held in 
Bolzano\footnote{http://ssu2015.inf.unibz.it/} (Italy) in 2015, and the second 
took place in Trondheim\footnote{https://iwssublog.wordpress.com/} (Norway) in 
2016. This paper provides a research agenda based on the activities carried out by the 
researchers in the network.

The rest of the paper is organized as follows. After we clarify the meaning of 
\textit{software startup} and what we know about software startups from prior research in 
the Background section, Section~\ref{agenda} introduces the research topics on 
software startups, organized under six main tracks that we have either investigated or 
envision investigating in the future. Wherever possible, each topic is 
illustrated and motivated by previous studies. Section~\ref{discussion} highlights 
the implications of these main tracks for future research. The paper concludes with 
Section~\ref{outlook}, which points out future actions that can establish and consolidate 
software startups as a research area.


\section{Background}

\subsection{What is a Software Startup?}
To understand software startups, we must first clarify what a startup is. According to 
Ries~\cite{ries_lean_2011}, a startup is a human institution 
designed to create a new product/service under conditions of extreme uncertainty. 
Similarly, Blank~\cite{Blank2005} describes a startup as a 
temporary organization that creates high-tech innovative products and has no prior 
operating history. These definitions distinguish startups from 
established organizations that have more 
resources and already command a mature market. In addition, Blank~\cite{Blank2005, 
Blank2012} defines a startup as a temporary organization that seeks a scalable, 
repeatable, and profitable business model, and therefore aims to grow. 
Blank's definition highlights the difference between a startup and a small 
business, which does not necessarily intend to grow, and consequently lacks a
scalable business model.

Even though sharing common characteristics with other types of startups, such as resource 
scarcity and a lack of operational history, software startups are often caught up in the 
wave of technological change frequently happening in software industry, such as new 
computing and network technologies, and an increasing variety of computing devices. They 
also need to use cutting-edge tools and techniques to develop innovative software 
products and services~\cite{sutton_role_2000}. All these make software startups 
challenging endeavours and meanwhile fascinating research phenomena for software 
engineering researchers and those from related disciplines.    

In 1994, Carmel first introduced the term \textit{software startup}, or, to be more 
precise, \textit{software 
package startup}, in SE literature~\cite{Carmel1994}. Carmel~\cite{Carmel1994} argued 
that software was increasingly becoming a fully realized product. Since then, other 
researchers have offered their own definitions of \textit{software startup}. 
Sutton~\cite{sutton_role_2000} considers software startups as organizations 
that are challenged by limited resources, immaturity, multiple influences, 
vibrant technologies, and turbulent markets. Hilmola et al.~\cite{Hilmola2003} claim that 
most software startups are product-oriented and develop cutting edge software products. 
Coleman and Connor~\cite{Coleman2008} describe software 
startups as unique companies that develop software through various processes and without a prescriptive methodology. 

Currently, there is no consensus on the definition of \textit{software startup}, 
even though many share an understanding that software startups deal with uncertain 
conditions, grow quickly, develop innovative products, and aim for scalability. Different 
definitions emphasize distinct aspects, and consequently
may have varying implications for how studies that adopt them should be designed, e.g., 
who qualifies as study subjects, or which factor is 
worth exploring. For this reason, despite the lack of a single 
agreed-upon definition of \textit{software startup}, it is important and 
recommended that researchers provide an explicit characterization of the 
software startups they study in their work. The research track in 
Section~\ref{sec:context} is dedicated to develop a software startup context model that 
would allow for such a characterization.

\subsection{What are the Major Challenges of Software Startups?}
Software startups are challenging endeavours, due to their nature as newly created 
companies operating in uncertain markets and working with cutting edge technology. 
Giardino et al.~\cite{Giardino2014b} highlight software startups' main challenges as: 
their lack of resources, that they are highly reactive, that they are by definition a new 
company, that they are comprised of small teams with little experience, their reliance on 
a single product and innovation, and their conditions of uncertainty, rapid evolution, 
time pressure, third-party dependency, high risk, and dependency (they are not 
self-sustained). Further, Giardino et al.~\cite{giardino_key_2015} apply the
MacMillan et al.~\cite{Macmillan1987} framework in the software startup context,
categorizing the key challenges faced by early stage software
startups into four holistic dimensions: product, finance, 
market, and team. The findings of Giardino et al.~\cite{giardino_key_2015} reveal that 
thriving in technological uncertainty and acquiring the first paying customer are
the top key challenges faced by many startups. In another study, 
Giardino et al.~\cite{Giardino2014} discover that inconsistency between 
managerial strategies and execution could lead to startup failure.  

Although research exists on the challenges software startups face, there is no study 
dedicated to their success factors. Block and Macmillan's~\cite{Block1985} study 
highlights the success factors for any new business, including generating ideas to 
complete product testing, completing a prototype, and consistently re-designing or
making amendments. Researchers have yet to explore these general factors' applicability 
to the specific software startup context.

\subsection{What do We Know about Software Engineering in Software Startups?}
Software development comprises a software startup's core activity. However, some initial 
research studies report a lack of software engineering activities in software startups. A 
systematic mapping study conducted by Paternoster et al.~\cite{paternoster_software_2014} 
allows us to start understanding how software startups perform software development. The 
study reveals that software requirements are often market driven and are not very well 
documented. Software development practices are only partially adopted; instead, pair 
programming and code refactoring sessions supported by ad-hoc code metrics are common 
practices. Testing is sometimes outsourced or conducted through customer acceptance and 
focus groups, and team members are empowered and encouraged to adapt to several roles. 
Similarly, Giardino et al.~\cite{Giardino2014b} highlight the most common development 
practices that have been used in software startup companies, such as: using well-known 
frameworks to quickly change the product according to market needs, evolutionary 
prototyping and experimenting via existing components, ongoing customer acceptance 
through early adopters’ focus groups, continuous value delivery, focusing on core 
functionalities that engage paying customers, empowerment of teams to influence final 
outcomes, employing metrics to quickly learn from consumers’ feedback and demand, and 
engaging easy-to-implement tools to facilitate product development. 

Although a few studies provide snapshots of software engineering practices in software 
startups~\cite{klotins_software_2016, giardino_software_2016}, the state of the art 
presented in literature is not enough to base an understanding of how software 
engineering practices could help software startups. 
Researchers must build a more comprehensive, empirical knowledge base in order to support 
forthcoming software startups. 
The research agenda presented in this paper intends to inspire and facilitate  
researchers interested in software startup related topics to start 
building such knowledge base.

\section{Research Agenda}\label{agenda}
\begin{figure}
	\centering
	\includegraphics[scale=0.5]{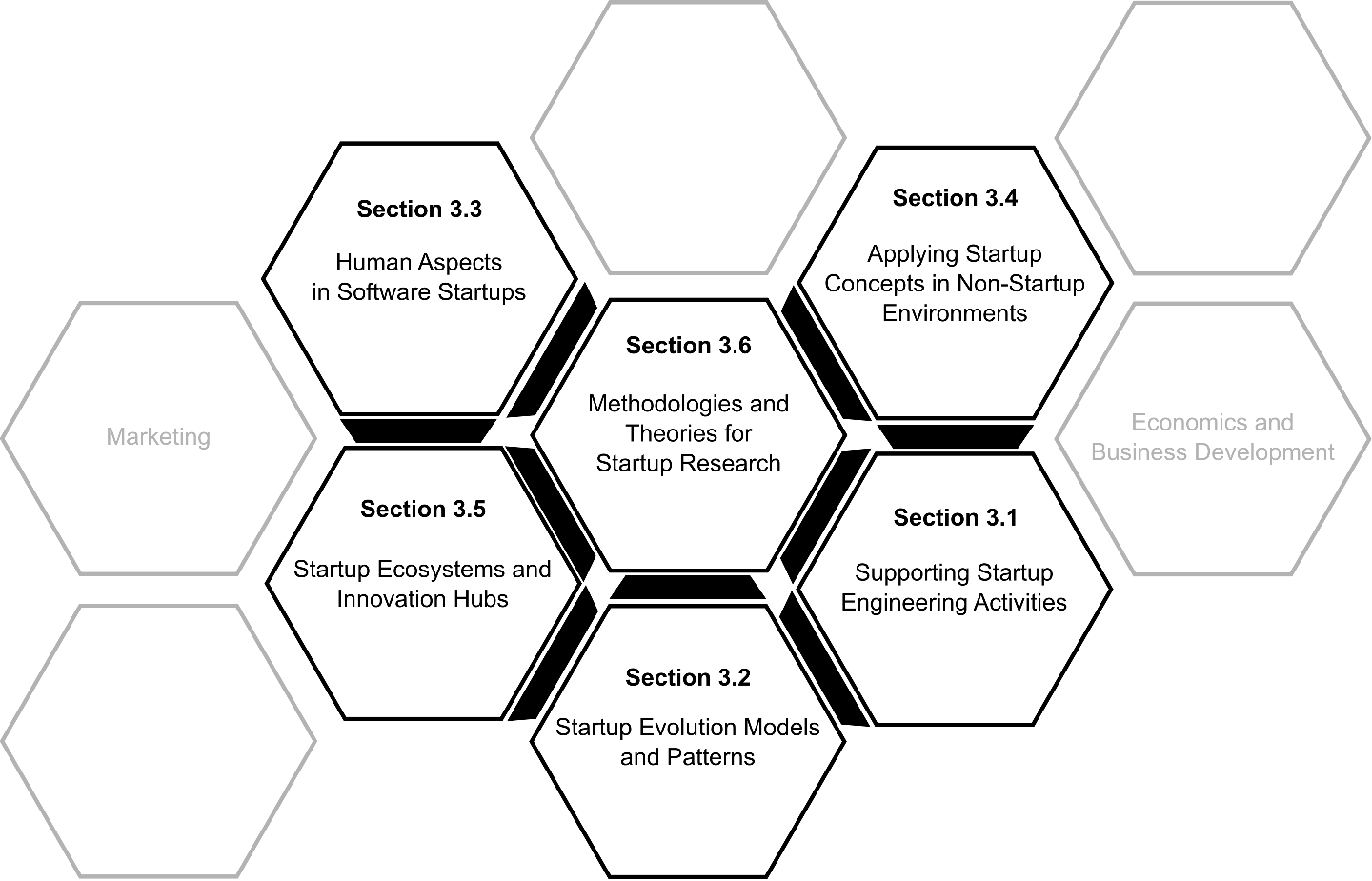}
	\caption{Overview of the Software Startup Research Agenda}
	\label{fig:ssra} 
\end{figure}
The Software Startup Research Agenda, initialized in June 2015, was developed by a 
network of researchers interested in studying the startup phenomenon from different 
angles and perspectives. This variety of research interests not only opens up new avenues 
for collaboration, but also sheds light on the complexity of the studied 
phenomenon. Initially, ten researchers created a mind map of different research areas, 
aiming to provide an overview of software startup research areas and how they connect to 
each other. Over a period of six months, more researchers joined the network,  added their 
research tracks, and continuously expanded the map. A working session with 
twenty researchers at the 1st workshop on software startup research in December 2015 was 
devoted at discussing the identified areas and finding potential interest overlaps among 
the participants. After this meeting, the authors of this paper prepared 
eighteen research track descriptions according to the following pattern: background of the 
area, motivation and relevance for software engineering in startups, research questions, 
potential impact of answering these research questions on practice and research, potential 
research methodologies that can be employed to answer the proposed research questions, and 
related past or ongoing work. Most of the authors interacted in the past or are currently 
active as advisory board members, mentors, founders or team members of software startups. 

The leading authors of this paper grouped the eighteen research 
tracks into six major clusters, based on the thematic similarities and differences of the 
tracks. While this grouping is one of the several possible ways to create the clusters, 
it served the purpose to ease the presentation and discussion of the 
research agenda, shown in Figure~\ref{fig:ssra}. Supporting Startup Engineering 
Activities (Section~\ref{sec:cluster1}) encompasses research foci that address specific 
software engineering challenges encountered by startup companies. Startup Evolution 
Models and Patterns (Section~\ref{sec:cluster2}) focuses on the progression of startups 
over time, trying to understand the underlying mechanics that drive a company 
towards success or failure. Human Aspects in Software Startups 
(Section~\ref{sec:cluster3}) covers research tracks that 
investigate factors related to the actors involved in startups. The research on 
Applying Startup Concepts in Non-Startup Environments 
(Section~\ref{sec:cluster4}) seeks to strengthen innovation by extracting successful 
software startup practices and integrating them in traditional environments. 
Startup Ecosystems and Innovation Hubs (Section~\ref{sec:cluster5}), on the 
other hand, investigates whether and how a thriving environment for software startups 
can be designed. Finally, all of these areas are connected by research tracks that 
develop methodologies and theories for software startup research 
(Section~\ref{sec:cluster6}).

Figure~\ref{fig:ssra}'s illustration of the research agenda includes reference to
research areas outside this paper's current scope. Marketing and 
Business and Economic Development are directions that are likely relevant for the 
performance of software startups. These and other areas may be added to the research 
agenda in later editions when more evidence exists regarding whether and how they 
interact with software startup engineering, i.e. the ``use of scientific, engineering, 
managerial and systematic approaches with the aim of successfully developing software 
systems in startup companies''~\cite{giardino_software_2016}.

\subsection{Supporting Startup Engineering Activities}\label{sec:cluster1}
The research tracks in this cluster share the theme of studying, identifying, 
transferring, and evaluating processes, methods, framework, models, and tools 
aimed at supporting software startup engineering activities. 

\subsubsection{The context of software intensive product engineering in 
startups}\label{sec:context}
Rapid development technologies have enabled small companies to quickly build and 
launch software\-intensive products with few resources. Many of these attempts 
fail due to market conditions, team breakup, depletion of resources, or a 
bad product idea. However, the role of software engineering 
practices in startups and their impact on product success has not yet been explored 
in depth. Inadequacies in applying engineering practices could be a significant 
contributing factor to startup failure.

Studies show that startups use ad-hoc engineering practices or attempt 
to adopt practices from agile approaches~\cite{yau_is_2013, 
klotins_software_2015}. 
However, such practices often focus on issues present 
in larger companies and neglect startup-specific challenges. For example, Yau and 
Murphy~\cite{yau_is_2013} report that test-driven development and pair programming 
provide increased software quality at an expense of cost and time. Also keeping to a 
strict backlog may hinder innovation. Since neglecting engineering challenges can lead to 
sub-optimal product quality and generate waste, engineering practices specific to the 
startup context are needed. The overarching questions in this research track are:
\begin{itemize}
	\item RQ1: To what degree is the actual engineering a critical success factor 
	for startups?
	\item RQ2: How can the startup context be defined such that informed decisions 
	on engineering choices can be made?
	\item RQ3: What engineering practices, processes and methods/models are used 
	today, and do they work in a startup context?
\end{itemize}

An answer to RQ1 could help practitioners to decide on what activities to focus on and 
prioritize allocation of resources. 
Several studies, e.g Paternoster et al.~\cite{paternoster_software_2014}, 
Giardino et al.~\cite{giardino_key_2015} and Sutton~\cite{sutton_role_2000}, 
emphasize the differences between established companies and startups, 
noting that startups are defined by limited resources and dynamic technologies. However, 
these characterizations are not granular enough to support a comparison of 
engineering contexts in different companies, making the transfer of practices from 
company to company difficult~\cite{petersen_context_2009}. Thus, understanding the 
engineering context of startups (RQ2) is an important milestone in developing startup 
context specific engineering practices (RQ3). While there exists work that provides 
systematic context classifications for the field of software engineering in 
general~\cite{petersen_context_2009,clarke_situational_2012, dyba_a_what_2012, 
kirk_investigating_2014, kirk_categorising_2014}, these models are not 
validated and adapted for use within startups. The work in this research 
track aims to develop such a software startup context model by analysing data from 
startup experience reports~\cite{klotins_software_2016}.
Provided that engineering contexts among startups and established companies can 
be compared at a fine level of detail, the context model can be used to identify 
candidate practices. Moreover, researchers can develop decision support by mapping 
specific challenges with useful practices, thereby validating the model and helping 
practitioners select a set engineering practices for their specific context and set of 
challenges.

\subsubsection{Technical debt management}
The software market changes rapidly. As discussed by Feng et 
al.~\cite{Feng2012929}, in fast changing environments, the product management 
focus evolves from the more traditional cost or quality orientation to a 
time orientation. New product development speed is increasingly important for 
organizations, and a commonly shared belief is that time-to-market of new 
products can build a competitive advantage~\cite{Feng2012929}. In the software 
startup context, it may be vital to be the first to market in order to obtain 
customers. Since software startups also lack resources, quality assurance is often 
largely absent~\cite{paternoster_software_2014}. However, long-term problems 
will only be relevant if the product obtains customers in the short 
term~\cite{tom_exploration_2013}. 
This short-term vision may produce software code that is low-quality and difficult to 
change, compelling the company to invest all of its efforts into keeping the system 
running, rather than increasing its value by adding new 
capabilities~\cite{tom_exploration_2013}. Scaling-up the system may become an obstacle, 
which will prevent the company from gaining new customers. Finding a viable trade-off 
between time-to-market demands and evolution needs is thus vital for software 
startups. 

One promising approach to performing such a trade-off is technical debt management. 
Technical debt management consists of identifying 
the sources of extra costs in software maintenance and analysing when it is 
profitable to invest effort into improving a software
system~\cite{tom_exploration_2013}.  
Hence, technical debt management could assist startups in making decisions on 
when and what to focus effort on in product development. Technical debt 
management entails identifying the technical debt sources, the impact 
estimation of the problems detected, and the decision process on whether it 
is profitable to invest effort in solving the detected sources of technical 
debt~\cite{fernandez_2015, Li2015}. Only those 
sources of technical debt that provide return on investment should be resolved.
More importantly, technical debt should be managed during project 
development~\cite{Lim2012} in order to control the internal quality of the developed 
software. Several research questions need to be answered to successfully manage technical 
debt in this way:
\begin{itemize}
	\item RQ1: What kind of evolution problems are relevant in the software startup 
	context? How can we identify them?
	\item RQ2: How can we prioritize the possible improvements/changes in the context of 
	software startups?
	\item RQ3. What factors beyond time-to-market and resource availability must be 
	considered 
	in trade-offs?
	\item  RQ4: How can we make decisions about when to implement the 
	improvements/changes within the software startup roadmap?
	\item RQ5: How can we provide agility to technical debt management, necessary in an 
	environment plenty of uncertainty and changes?
\end{itemize}
\medskip
Answering these questions will impact on both practitioners and researchers focused on 
software startups. Practitioners will be able to make better decisions considering the 
characteristics of the current software product implementation. The current  
implementation could make it impossible to reach a deadline (time to market), because of 
the complexity of the changes to perform to implement a new feature, assuming a given 
amount (and qualifications) of effort to be deployed. Furthermore, it will be also 
possible to decide between two alternative implementations, with different costs, but 
also with different potential for the future, assuming that the ``future'' has been 
previously outlined. For researchers, answering these questions could help clarify the 
role of design decisions in software development in the context of a software product 
roadmap, similarly to what happens in other engineering disciplines.

Technical debt is context dependent since quality tradeoffs are context 
dependent~\cite{shull_technical_2013}. While technical debt is as important to software 
startups as it is to mature companies, the kind of decisions to take and the consequences 
of making the wrong decisions are not the same, justifying research on technical debt 
specifically in software startups.

In general, there is a lack of specific studies on technical debt management in 
software startups, and current literature reviews on technical debt management do 
not address this topic~\cite{fernandez_2015,Li2015}. Moreover, there are 
several specific challenges to managing technical debt that are of special 
relevance for software startups. For one, very few studies 
address how to prioritize improvements to solve technical debt problems, 
especially for commercial software development~\cite{Li2015}. In addition, technical debt 
management literature often refers to time-to-market, but 
very few studies actually address it~\cite{fernandez_2015}, perhaps because it is a topic 
that straddles engineering and economics.

\subsubsection{Software product innovation assessment}\label{sec:innovation}
Startup companies strive to create innovative products. For firms in general, 
and software startups in particular, it is critical to know as soon as possible 
if a product aligns with the market, or whether they can increase their chances to lead 
the market and recruit the highest 
possible number of customers~\cite{Johne1988114}. 

The need to invest in infrastructures to measure the impact of innovation in 
software was highlighted by OECD~\cite{OECDInnovation}, and more recently by 
Edison et al.~\cite{edison_towards_2013}. These measures will enable companies 
to assess the impact of innovation factors and achieve the expected business goals, 
as well as to improve the understanding of success yield high returns on investments in 
the innovation process~\cite{OECDInnovation}. Product 
innovation assessment is thus very relevant for product developers, and especially for 
startups, which are more sensitive to market reactions. Product innovation 
assessment is complex, particularly for software 
products~\cite{eversheim_innovation_2008}.

Product innovation assessment is reported in literature as the combination 
of a number of multi-dimensional factors impacting the success or failure of a 
software product~\cite{Crossan_Apaydin_2009}. Factor's measures intend to 
engage people in the innovation process to think more deeply about factors 
affecting product innovation. Factors such as time-to-market, perceived 
value, technology route, incremental product, product liability, risk 
distribution, competitive environment, life cycle of product, or strength of 
market could be grouped into dimensions like market, organization, environment, 
or any other terms of impact on the market and business drivers~\cite{618169}. 
These factors can act as innovation enablers or blockers~\cite{cooper1999}. 

Since these factors are not always independent, it is critical 
to identify the existing dependencies and gain a better understanding of 
each factor's impact. It would be necessary to relate these 
factors to characteristics specific to software products, such as, but not 
limited to, software quality attributes proposed by ISO/IEC~\cite{isoiec25000}.

There is a lack of specific literature on \emph{software} product innovation 
assessment; most of the past research refers to products in general, and not 
specifically to software products~\cite{yague_analyzing_2014, edison_towards_2013}, 
leading to the following research questions:
\begin{itemize}
	\item RQ1: What should be the components of a software product innovation 
	assessment/estimation model? 
	\item RQ2: What factors can help measure innovation from a software product and a 
	market perspective?
	\item RQ3: To what extent are factors that can help measure innovation dependent on 
	the	software product and the market perspective?
	\item RQ4: What is the relation between software product innovation factors and 
	quality factors?
	\item RQ5: What kind of tools for software product innovation estimation could 
	support software startups in decision making?
\end{itemize}
\medskip
While innovation has been widely studied from the process perspective, the product 
perspective, by nature, has been addressed mainly from the viewpoint of specific products 
and industries. However, software products are different compared to other kinds of 
products~\cite{pikkarainen_art_2011} and innovations in the software industry happen fast.
Hence, answers to RQ1-RQ4 would provide a fundamental understanding on software product 
innovation assessment and be beneficial for both researchers and practitioners.
Software startups need to be fast and spend resources in an efficient way. Therefore, to 
be able to estimate existing products or design new products, considering those 
characteristics that experience shows that are relevant from an innovation point of view, 
can be essential for software startups to develop successful products (RQ5).

\subsubsection{Empirical prototype engineering}\label{sec:prototype}
Startups often start with a prototype, which serves as a form to validate either a new 
technology or knowledge about targeted customers~\cite{paternoster_software_2014}. 
Traditionally, prototyping implies a quick and economic approach to determining final 
products~\cite{lichter_prototyping_1993, beaudouin-lafon_prototyping_2002, 
sommerville_software_2010}. Defined as a concrete representation of part or all of an 
interactive system, prototypes has been intensively researched and used in Software 
Engineering, with well-developed taxonomies, such as horizontal and vertical, 
low-fidelity and high-fidelity prototypes~\cite{sommerville_software_2010}. The strategy of 
developing a prototype can greatly vary due to a great variety of prototype types, their 
development efforts and value they can produce.

While much about prototyping techniques can be learnt from the SE body of knowledge, the 
discussion about prototyping in the context of business development process is rare. 
Recent work on startup methodologies, such as Lean Startup~\cite{ries_lean_2011} and 
Design Thinking~\cite{brown_change_2009} emphasizes the adoption of prototypes to 
increase chances of success through validated learning. Alternatively, startup prototypes 
need to be developed to satisfactorily serve their purposes, i.e. technical feasibility 
test, demonstration to early customers, and fund raising. We argue that the prevalent 
Software Engineering practices used by startups to develop their first product 
inefficiently integrate into startups’ dynamic contexts. Hence we call for research in 
understanding the development and usage of prototypes in startup contexts:
 
\begin{itemize}
	\item RQ1: How can prototyping be used to maximize learning experience? 
	\item RQ2: How can prototyping be used for optimization?
	\item RQ3: How can prototyping be used to support communication with external 
		stakeholders?
	\item RQ4: How do prototypes evolve under the multiple influences of startups' 
	stakeholders?     	
\end{itemize}
\medskip
Early stage startups are lacking actionable guidelines for making effective prototypes 
that can serve multiple purposes. We believe that many startups will economically and 
strategically benefit by having proper practices in prototyping, such as technology 
evaluation (RQ1), strategic planning (RQ2) and customer involvement (RQ3).

To understand prototype development and its usage in startups, i.e. answering the first 
three research questions, exploratory case studies can be conducted. Cases would be 
selected to cover different types of startup prototypes at different phase of startup 
progress. A large-scale survey can be used to understand the prototype usage patterns, 
i.e. answering RQ4. 

Despite an increasing body of knowledge on software 
startups~\cite{paternoster_software_2014}, empirical research on prototyping processes 
and practices are rare. A few studies have investigated the adoption of software 
prototypes in combination with Design Thinking~\cite{Efeoglu2013} and proposed 
prototyping techniques~\cite{Newman2015,Efeoglu2013,Grevet2015}. However, these studies 
rely on a very limited number of cases. Moreover, different constraints on prototyping 
decisions are often neglected. Future work can address antecedence factors, i.e. the 
involvement of lead-users, available human resources, and technological push, and how 
they impact prototyping strategies and usages in different startup 
contexts~\cite{NguyenDuc2016}.

\subsubsection{Risk Management Tools for Software Startups}
The management of risk, namely the risk of failing to meet one's goals 
within given constraints in budget and/or time, is of paramount importance in 
every human activity. In the context of software startups, risk management looks 
unconventional, because startups naturally involve a much higher risk 
than traditional businesses. Yet, perhaps even more so than in traditional 
contexts, evaluating and managing risk in the software startup context might be a key 
factor for success.

Risk factors can be identified as a check-list of the incidents or 
challenges to face. Each of them could be categorized and prioritized according to its 
probability and the impact level of its consequences. This research track aims to study, 
model, and quantify various aspects related to risk management in software startups, with 
the goal of providing tools, based on process simulation, that control risk. Being able 
to efficiently model and simulate the startup process and its dynamics, would support 
startups in timely decision making. While numerous other approaches to risk control 
exist~\cite{raz_use_2001}, we have found in our previous 
work~\cite{cocco_simulating_2011,concas_simulation_2013} that process simulations can be 
effective in risk management. Therefore, the overarching questions in this research track 
are:  

\begin{itemize}
	\item RQ1: To what extent do software startups explicitly manage risk? 
	\item RQ2: To what degree is it feasible to model software development processes in 
		startups?
	\item RQ3: To what extent can these models be used to quantify the risk of exceeding 
		project budget or time?
	\item RQ4: What systematic ways exist to understand when to pivot or 
	persevere~\cite{ries_lean_2011}, and what might be the cost of a wrong or untimely 
	decision? 
\end{itemize}
\medskip
Following our previous experiences in software process modelling and simulation, to gain 
a better understanding is necessary to identify and analyse significant activities, not 
limited to the software development phase, of a software startup (RQ1). 
This is necessary to be able to identify the critical aspects of startup development 
risks that are suitable for simulation. In our previous work we studied the 
application of Event-Driven models and/or System Dynamics to the software development 
processes. From this work we know that it is possible analyse project variations in time 
and budget with a Monte Carlo approach, by performing several simulations of the same 
project, varying the unknown parameters according to given distributions, and calculating 
the resulting distributions of cost and time of the simulated projects. Such analysis 
allows one to compute the Value At Risk (VAR) of these quantities, at given VAR levels. 
While Cocco et al.~\cite{cocco_simulating_2011} and Concas et 
al.~\cite{concas_simulation_2013} provide exemplar studies of the application of these 
techniques in mature (agile) software development contexts, the question is whether such 
an approach is suitable and beneficial for software startups, and under what conditions 
(RQ2). 
By simulating the evolution of a startup as a process, we might be able to make 
predictions on its future development. Such predictions, or a result that can be rapidly 
be drawn from simulations, might be crucial for startups to understand which decisions 
are less costly and/or risky (RQ3).  
This is particularly true for decisions related to fields such as market strategies, team 
management, financial issues or product development (RQ4).

\subsubsection{Startup support tools}
Support tools can help software startups get their business off the 
ground with less pain and more guidance. These tools generally embed crucial 
knowledge regarding startup processes and activities. A plethora of tools (mostly 
software tools) exist for meeting the different needs of entrepreneurs 
and supporting various startup activities. For example, the 
web-page\footnote{http://steveblank.com/tools-and-blogs-for-entrepreneurs/} by 
Steve Blank, a renowned entrepreneurship educator, author, and researcher from 
Stanford University, contains a list of more than 1000 tools. Well-designed 
portals such as Startupstash.com ease access to these supporting tools.

However, due to the lack of time, resources, and/or necessary knowledge, 
entrepreneurs cannot easily find the tools that best suit their needs, or 
cannot effectively utilize these tools to their potential. Existing studies provide 
limited insights on how entrepreneurial teams could find, use and benefit from support 
tools. Hence, the overarching questions in this research track are:
\begin{itemize}
	\item RQ1: What are the needs of software startups that can be supported by 
	software tools?
	\item RQ2: What are the tools that support different startup activities?
	\item RQ3: How can support tools be evaluated with respect to their 
	efficiency, effectiveness, and return-on-investment?
	\item RQ4: How can support tools be effectively recommended to entrepreneurs and used 
	by them?
\end{itemize}
\medskip
RQ1 and RQ2 are targeted at identifying a match between the needs of software startups 
and the available tool support. To enable robust recommendations, both the individual 
startups and the software tools need to be objectively characterized allowing for their 
evaluation w.r.t. certain quality criteria (RQ3). There are potential synergies with 
the research track looking at the context characterization of software startups 
(Section~\ref{sec:context}). Answers to these research questions can be also valuable 
input for software tool vendors to develop the right tools that are needed by startups. 
In addition, the findings can be useful for future studies that develop proof-of-concept 
prototypes to support startup activities.

To investigate the proposed questions, various research methods can be applied, including 
survey of software startups regarding their needs and usage of support tools, in-depth 
case study of adoption and use of support tools, and design science approach to develop 
recommender systems of support tools (RQ4).

Research on tooling aspects in the software startup context is scarce. 
Edison et al.~\cite{edison_towards_2015} argue that, despite the 
fact that different startup supporting tools have been developed and published 
over the Internet, new entrepreneurs might not have sufficient knowledge of 
what tools they need when compared to experienced entrepreneurs. In addition, not all 
tools will help entrepreneurs in certain tasks or situations. Entrepreneurs' experiences 
using the tools can serve as the basis for evaluating and recommending appropriate tools. 
Besides suggesting a new categorization of existing startup support tools, Edison et 
al.~\cite{edison_towards_2015} propose a new design of a tool portal that will 
incorporate new ways to recommend tools to entrepreneurs, especially to those who engage 
for the first time in a software startup endeavour.

\subsubsection{Supporting software testing}\label{sec:testing}
Testing software is costly and often compromised in 
startups~\cite{zettell_lipe:_2001}, as it is challenging for startups to 
fulfil customer needs on time, while simultaneously delivering a high quality product. 
In many software startups there is a common slogan that says ``done is better than 
perfect'', which indicates a general tendency toward a lack of testing and quality 
assurance activities~\cite{kelly_lessons_2012}. However, it is sometimes also observed 
that startups do not know how and what to test; they lack 
expertise to test requirements as they do not have knowledge about their 
customers and users~\cite{kelly_lessons_2012}. Therefore 
considering testing in software startups poses the following research questions:
\begin{itemize}
	\item RQ1: To what extent does software testing in startup companies differ from 
	traditional companies?
	\item RQ2: To what extent does testing evolve over time in software startup companies?
	\item RQ3: What is an optimal balance between cost/time spent on testing and 
	development activities?
	\item RQ4: How can a software startup leverage customers/users for testing?
\end{itemize}
\medskip
Answering RQ1 would provide insights on the aspects that differentiate the software 
testing process in startups from mature companies. For example, integration testing is 
likely very important for startups due to the fast paced product development. At the same 
time however, startups tend to work with cutting edge technologies, requiring a robust 
and flexible test integration platform. Connected to this is the question whether testing 
needs change over time, while the software startup matures. Answers to RQ2 and RQ3 
would be particularly valuable for practitioners who could then better allocate 
resources. Users of software could be used for different testing purposes. On one hand, 
users provide valuable feedback in testing assumptions on customers needs. On the other 
hand, early adopters that are more robust towards deficiencies can help to improve 
product quality before targeting a larger market. Answers to RQ4 would provide strategies 
to harvest these resources. 

In order to answer these research questions, various empirical research methods could be 
utilized. The studies would be devised in a way that ``contrasting results but for 
anticipatable reasons'' could be expected~\cite{yin_case_2003}, i.e. different software 
startup companies would be taken into account to acquire a broad view of testing in 
software startups.

To the best of our knowledge, software testing in software startups has been scarcely 
researched. Paternoster et al.~\cite{paternoster_software_2014} highlighted the quality 
assurance activities in software startups in their mapping study. They found that it is 
important to provide software startups effective and efficient testing 
strategies to develop, execute, and maintain tests. In addition, they highlighted 
the importance of more research to develop practical, commercial testing solutions 
for startups.

\subsubsection{User experience}
User experience (UX) is described as ``a person's perceptions and responses 
that result from the use or anticipated use of a product, system or 
service''~\cite{iso_2010}. Good UX can be seen as providing value to users, as well 
as creating a competitive advantage. UX is important for 
software startups from their earliest stages. Firstly, human-centred design methods such 
as user research and user testing can help startups better understand how they can 
provide value to users and customers, as well as what features and qualities need testing 
for users to be satisfied with their product. Combined with business 
strategy, this human-centred approach helps startups move towards successful, 
sustainable business creation. Secondly, providing an initially strong UX in the first 
product versions can create positive word of mouth~\cite{fuller_user_2013}, as well as 
keep users interested in the product for a longer time~\cite{hokkanen_ux_2015}. 
Genuine interest from users for the product idea while the product is still a 
prototype helps gain meaningful feedback~\cite{hokkanen_ux_2015}. 
Compared to more established businesses, software startups may pivot resulting in new 
target markets and user groups. This means efforts put into designing UX need to be 
faster and less resource consuming. Furthermore, failing to deliver satisfying UX can be 
fatal to small startups that can not cover the costs of redesigning.  The overarching 
questions in this research track are:
\begin{itemize}
	\item RQ1: What useful methods and practices exist for creating UX in startups?
	\item RQ2: What is UX's role during different phases of a startup's life-cycle?
	\item RQ3: To what extent are UX and business models connected in customer value 
	creation?
\end{itemize}
\medskip
An answer to RQ1 can provide software startups methods for developing strong UX in the 
first product versions which can keep users interested in the product for a longer 
time~\cite{hokkanen_ux_2015}. Genuine interest from users for the product idea while 
the product is still a prototype helps to gain meaningful 
feedback~\cite{hokkanen_ux_2015}. For business creation, understanding the value of UX 
for startups (RQ2) helps assigning enough resources for creation of UX while not wasting 
resources where there is no value to be gained (RQ3). 

Research on startups and UX has been very limited. Some case studies report UX's role in 
building successful startups~\cite{may_applying_2012, 
taipale_huitalestory_2010}. Practices and methods for UX work in startups have 
been reported in~\cite{hokkanen_ux_2015, hokkanen_early_2015, 
hokkanen_three_2015}. A framework for creating strong early UX
was presented by Hokkanen et al.~\cite{hokkanen_ux_2016}. These provide 
some results on feasible and beneficial UX development in startups, but more 
generalizable results are needed.

\subsection{Startup Evolution Models and Patterns}\label{sec:cluster2}
The research tracks in this cluster share the theme of studying, identifying, 
and differentiating the transformation of startups in different stages. This also 
includes studies about different business and technical decision-making 
practices.

\subsubsection{Pivots in software startups}
It is very difficult for software startups to understand from start what are the real 
problems to solve and what are the right software solutions and suitable business models. 
This is evidenced by the fact that many successful software startups are different from 
what they started with. For example, Flickr, a popular online photo sharing web 
application, originally was a multiplayer online role playing game~\cite{Nazar2013}. 
Twitter, a famous microblogging application, was born from a failed attempt to offer 
personal podcast service~\cite{Nazar2013}. 

Due to their dynamic nature, software startups must constantly make crucial 
decisions on whether to change directions or stay on the chosen 
course. These decisions are known as \textit{pivot} or \textit{persevere} 
in the terms of Lean 
Startup~\cite{ries_lean_2011}. A pivot is a strategic decision used to test 
fundamental hypothesis about a product, market, or the engine of 
growth~\cite{ries_lean_2011}. Software startups develop technology intensive products in 
nature. Due to this, these are more prone to the rapidly changing technology causing 
pivots. Similarly, certain types of pivots are more relevant to software startups e.g. 
zoom in pivot: a pivot where one feature of a product become the whole product as in the 
case of Flickr. Pivot is closely linked to validated learning, another key concept from 
Lean Startup. The process to test a business hypothesis and measure it to validate its 
effect is called validated 
learning~\cite{ries_lean_2011}, whereas pivot is often the outcome of validated 
learning. A recent study~\cite{Giardino2014} reveals that startups often 
neglect the validated learning process, and neglect pivoting when they need to, 
which leads to failure. This shows the importance of pivoting for a startup to 
survive, grow, and eventually attain a sustainable business model. 
In order to better understand and explore the pivoting process in the software startup 
context, the following fundamental research questions can be formed:

\begin{itemize}
	\item RQ1: To what extent is pivoting crucial for software startups?
	\item RQ2: How do software startups pivot during the entrepreneurial/startup process?
	\item RQ3: What are the existing process/strategies/methods to make a pivoting 
	decision in a startup context?
	\item RQ4: How do pivots occur during different product development and 
	customer development life cycles?
\end{itemize}
\medskip
Answering RQ1-RQ2 is necessary to understand pivoting in the context of software 
startups, building a fundamental framework on reasons for pivoting and their types. 
RQ3-RQ4, on the other hand, are targeted at understanding pivoting decisions and 
mechanisms.
The overall contribution of answering the stated research questions has implications for 
both researchers and practitioners. The answers would provide an empirically validated 
conceptual and theoretical basis for the researchers to conduct further studies regarding 
the pivot phenomenon. For the practitioners, it would help them to make informed decision 
regarding when and how to pivot in order to increase the chances of success.

Due to the nascent nature of software startup research area, exploratory cases studies is 
a suitable approach to answer the research questions. Followed by the case studies, 
quantitative surveys can also be conducted to further generalize the results regarding 
pivoting in software startups.

Recently, there were some studies conducted on pivots in software startups. A study by 
Van der Van and Bosch~\cite{VanderVen2013} compares pivoting decisions with software 
architecture decisions. Another study by Terho et al.~\cite{Terho2015} describes how 
different types of pivots may change business hypothesis on lean canvass model. However, 
these studies lack the sufficient detail to understand different types of pivots and 
the factors triggering pivots. A study by Bajwa et al.~\cite{ShahidBajwa2016}, presents 
an initial understanding of different types of pivots occurred at different software 
development stages, however it lacks the deeper understanding of the pivoting decision 
that can only be achieved by a longitudinal study.

\subsubsection{Determination of Software Startup Survival Capability through Business Plans}\label{sub:survival}
Software startups are highly specialized from a technological point of view. Focusing on 
the economic exploitation of technological innovations~\cite{lofsten_science_2002}, they 
belong to the group of new technology-based firms. Literature suggests that one 
of their major challenges is the transformation of technological know-how into marketable 
products~\cite{gans_product_2003,brem_integration_2009}. New technology-based firms often struggle with 
unlocking the product-market fit~\cite{maurya_running_2012} and commercializing their 
technological products~\cite{gans_product_2003}. Applying a resource-based view does thus 
not suffice for explaining survival and growth of software 
startups~\cite{klyver_resource_2013, levie_terminal_2010}: a crucial success factor is 
the ability of new technology-based firms to understand and interact with the market environment to position 
their products accordingly~\cite{giones_strategic_2015, clarysse_explaining_2011}.

Particularly in early lifecycle stages, new technology-based firms need to build network relations with the 
market. Network theory literature suggests that with increasing network maturity, the 
chances for survival and growth increase~\cite{newbert_supporter_2010, 
semrau_networking_2012, 
witt_entrepreneurs_2004}. The ability to transform resources in response to triggers 
resulting from market interactions can be described as a dynamic 
capability~\cite{eisenhardt_dynamic_2000, teece_dynamic_1997, 
newbert_looking_2008, lichtenstein_measuring_2006} which helps software startups 
commercialize their products. This transformation process captures the evolution of new technology-based firms 
in their early-stages. Current research is based on the construct of “venture emergence”, which provides a 
perspective on the evolutionary change process of new technology-based firms~\cite{giones_strategic_2015, 
brush_properties_2008}. Venture emergence reflects the interaction process with agents and their 
environments ~\cite{katz_properties_1988}. 
Business plans of new technology-based firms are used as the artefact for measuring the status of venture 
emergence. They 
contain descriptions of transaction relations~\cite{honig_institutional_2004, 
kirsh_firm_2009, karlsson_judging_2009} new technology-based firms build in four market dimensions: customer, 
partner, investor, and human resources~\cite{konig_role_2015}. This research track 
intends to answer a number of research questions:
\begin{itemize}
	\item RQ1: How reliably can annotated transaction relations from business plan texts 
	determine the venture emergence status of technology-based startups?
	\item RQ2: To what extent are the number and strength ("level") of identified 
	transaction relationships useful as an indicator of survival capability? 
	\item RQ3: How can patterns of transaction relations be used as an indicator for 
	evaluating strengths and weaknesses of new technology-based firms, and thus be used to more effectively 
	direct support measures?
\end{itemize}
\medskip
While it is possible to measure the venture emergence status even in a software startup’s very early 
stages, the predictive strength of transaction relations needs to be evaluated (RQ1-RQ2). 
This use of network theory to operationalize the venture emergence construct is a new approach, which 
adds to network theory literature in the context of the survival of new technology-based firms. It further confirms 
the business plans of new technology-based firms as a valuable source of information on startup potential. Finally, 
the resource-based approach to explain venture survival is enriched by applying a 
process-oriented perspective: we analyse resource transformation, rather than only 
looking at the initial resource configuration (RQ3). Furthermore, the research can 
contribute to the effectiveness of the innovation system by investigating indicators that 
reveal strengths and weaknesses of new technology-based firms. These can be used to direct support measures to 
software startups more effectively.

To answer the stated research questions, one can use content 
analysis~\cite{dovan_reliability_1998, elo_qualitative_2008}, combining human and 
computer-based coding of business plans, to determine the number and strength of 
transaction relations~\cite{konig_agreement_2016, ungerer_measuring_2016}. 

Initial statistical tests that have been performed on a sample of 40 business plans of new technology-based firms  
confirm the relationship between the status of venture emergence of new technology-based firms and venture 
survival~\cite{ungerer_measuring_2016}. Earlier work led to the development of the 
concept for analysing early-stage startup networks and the relevance for 
survival~\cite{konig_role_2015}. Based on this concept, a coding method for transaction 
relations in business plans has been developed and validated with 120 business 
plans~\cite{konig_agreement_2016}.

\subsection{Cooperative and Human Aspects in Software Startups}\label{sec:cluster3}
The research tracks in this cluster address challenges and practices related to 
how people cooperate and work is software startups.
  
\subsubsection{Competencies and competency needs in software 
startups}\label{sec:competency}
Software startups set different competency requirements on their personnel than more 
established companies. The biggest differences occur in two phases of the evolution of 
startups which have an impact on the nature of software development and competence needs: 
(1) in the early stages of rapid software development when there is a lack of resources 
and immature competencies in many key areas, and (2) when the rapid business growth of 
successful startups requires management of a fast growing personnel and amount of 
software with limited management resources and competencies. In the early phases strong 
competition requires the software startup to innovate and react 
quickly~\cite{paternoster_software_2014}, and deployment of systematic software 
engineering processes is many times replaced by 
light-weight ad-hoc processes and methods~\cite{klotins_software_2015, 
paternoster_software_2014}. 
The nature of software makes it possible for successful startups to scale 
fast~\cite{paternoster_software_2014}. Rapid software-driven growth 
requires fast scaling of the software production, distribution, and maintenance. The 
required competences also quickly evolve when software development moves from rapid 
greenfield prototyping to professional software development and management. Mastering 
this demanding situation often requires a broad prior skill basis from the startup team, 
including an ability to adjust to changes, and learn quickly. 

Research on specific skills and competency needs in software startups broadens not only 
the knowledge on software startups themselves, but also broadens the knowledge on 
software engineering conducted under the challenging circumstances of startups. Focusing 
the research on the early stages and on the growth period of the software startups, when 
the challenges of the software startups are the greatest~\cite{giardino_key_2015, 
Giardino2014}, brings the most valuable knowledge to both academia and practitioners. 
Competency research also brings human factors into 
focus~\cite{marlow_human_2006,jack_small_2006}, and reinforces the results of existing 
software startup research towards a more comprehensive modelling and understanding.
The research questions for studies on competencies and competency needs in 
software startups include:
\begin{itemize}
	\item RQ1: Software startup challenges and competency needs \textemdash\/ what 
	software development knowledge and skills are needed to overcome the challenges?	
	\item RQ2: What are the competency needs specific for software startups compared to 
	the more established software companies?
	\item RQ3: How do the competency needs change over the evolution of software startups?
	\item RQ4: How do the competency needs map onto the roles and responsibilities of the 
	startup teams in software startups?
	\item RQ5: How can the growth of software startups be managed in terms of competency 
	needs for software development practices, processes and recruitment?
\end{itemize}
\medskip
Research on software startups, including research on competency needs, provides the 
research and development of software engineering with new knowledge and viewpoints on how 
to direct the work in order to best address the specific challenges of the software 
startups (RQ1). In particular, differences to mature software companies are interesting 
to study (RQ2) considering software startups evolve, if they survive, to established 
companies. Knowing how competency needs change might turn out as one key factor for this 
transition (RQ3). Theoretical models describing the evolution paths of software startups 
have been created~\cite{ries_lean_2011,Bosch2013}, but competency needs and how they map 
to roles and responsibilities have been to a large degree ignored (RQ4). Similarly, while 
software development work~\cite{paternoster_software_2014} and software engineering 
practices~\cite{klotins_software_2015} have also been studied, it is unclear how 
competency needs can be managed in growing software startups (RQ5).

\subsubsection{Teamwork in software startups}
The importance of human aspects in software development is increasingly recognized by 
software engineering researchers and practitioners. Teamwork effectiveness is crucial for 
the successes of any product development project~\cite{Dingsoyr2013}. A common definition 
of a team is ''a small number of people with complementary skills who are committed to a 
common purpose, set of performance goals, and approach for which they hold themselves 
mutually accountable''~\cite{Katzenbach1993}. A startup team is special in the wide range 
of variety, including both technicians and entrepreneurs. 

While an innovative idea is important for the formation of a startup, startup success or 
failure ultimately rests on the ability of the team to execute. Entrepreneurship research 
showed that over 80 percent of startups that survive longer than two years were founded 
by a group of two or more individuals~\cite{Oechslein2012}. The dynamic and intertwined 
startups activities require the close collaboration not only among startup team members, 
but also with external stakeholders, such as mentors and investors. Given the diversity 
in mindsets and skill sets among founders, it is essential that they can work well 
together along with the startup life-cycle. The movement with recent methodology in Lean 
startup introduces an opportunity to look at startup teams from various angles, i.e. 
pivoting, startup culture, team formation, and decision-making. The overarching questions 
in this research track are:

\begin{itemize}
	\item RQ1: Is there a common cultural / organizational / team characteristic 
	among successful software startups?
	\item RQ2: How can a software startup team effectively communicate with other 
	stakeholders, i.e. mentors and investors? 
	\item RQ3: How can a software startup manage team internal relationships?
	\item RQ4: What are the common patterns of competence growth among software startup 
	teams?
\end{itemize}
\medskip
Understanding software startup team behaviour to internal and external environments and 
relating them to startup success measures would help to identify characteristics and 
teamwork patterns of successful startups. Answering RQ1 would provide practitioners some 
guidance on how to form startup teams while answers to RQ2-RQ3 would provide an 
understanding how internal end external team dynamics work and can improved. An answer 
to RQ4 would also support the work in Section~\ref{sec:competency}, looking however 
specifically at competence growth patterns that could be valuable for practitioners when 
deciding on what to focus on in competence development.
Empirical studies, i.e. case studies, surveys and action research are all suitable to 
investigate the stated research questions. Among them, comparative case studies would be 
the first option to discover the difference in startup teamwork patterns.

There exists a large body of literature in business management, entrepreneurship, and small 
ventures about entrepreneurial teams’ characteristics and their relationship to startup 
outcomes~\cite{Oechslein2012, Kamm1990, Francis2000}. In Software Engineering, few empirical 
studies identified team factors in the failure of software startups. Giardino et al. found 
that building entrepreneurial teams is one of the 
key challenges for early-stage software startups from idea conceptualization to the first 
launch~\cite{giardino_key_2015}. Crowne et al. described issues with founder teamwork, 
team commitment and skill shortages~\cite{Crowne2002}. Ensley et al. investigated the relative influence of 
vertical versus shared leadership within new venture top management teams on the 
performance of startups~\cite{Ensley2006}. Other team dimensions are explored 
in the business and engineering management domain in specific geographies. E.g., 
Oechslein analysed influencing variables on the relational capital dimension trust within 
IT startup companies in China~\cite{Oechslein2012}. How generalizable these influencing 
variables to other geographies is yet to be seen.  

\subsection{Applying Startup Concepts in Non-Startup 
Contexts}\label{sec:cluster4}
One of the Lean Startup principles claims that entrepreneurs are everywhere, 
and that entrepreneurial spirits and approaches may be applied in any size 
company, in any sector or industry~\cite{ries_lean_2011}. On the other hand, 
established organizations face the challenge of innovation dilemma and inertia 
caused by the organization's stability and the maturity of 
markets~\cite{Cooper2011}. Therefore, applying startup concepts in non-startup 
contexts seems an promising avenue for established organizations to improve 
their innovation potential.
   
\subsubsection{Internal software startups in large software companies}
\label{inernalstartup}

The internal software startup concept has been promoted as a way to nurture product 
innovation in large companies. An internal software startup operates within the 
corporation and takes responsibility for everything from finding a business idea to 
developing a new product and introducing it to market~\cite{Bart1988}. Internal 
software startups can help established companies master the challenge of 
improving existing businesses, while simultaneously exploring new future 
business that sometimes can be very different from existing ones~\cite{Hill2014}.
Usually, this involves a conflict of interest in terms of learning 
modes~\cite{Jansen2009} or risk propensity~\cite{Nanda2013}, 
which can be prevented by establishing dual structures within the organization 
for implementing internal software startups~\cite{Lavie2010}. Compared to 
the traditional R\&D activities of larger companies, an internal software startup 
develops products or services faster~\cite{paternoster_software_2014} and 
with higher market orientation~\cite{Lerner2013corporate}. This helps 
established companies maintain their competitiveness in volatile 
markets~\cite{OReilly2013}. 

Besides the fact that the successful implementation 
of internal software startups faces various barriers, such as cultural 
conflicts~\cite{Garvin2006} or the fear of cannibalization of existing 
businesses~\cite{Edison2015a}, internal software startups can also benefit from being 
part of established companies. Shared resources, such as capital, human 
resources~\cite{Chen2009, COLEMAN2013}, and the access to 
the corporates’ internal and external network~\cite{Dess2003} are just some 
benefits.

Earlier research on analysing the results of startups' value creation cycle
has taken place in the context of the evolution of the 
enterprise~\cite{croll_lean_2013}. However, this occurs over too long of a time period 
to be useful for guiding software development. Measuring the cycle time of the software
engineering process to the completion of a software feature is also
insufficient. The Lean startup approach~\cite{ries_lean_2011} has been 
commonly adopted to new business creation in software intensive ventures. They 
use the learning loop to discover the customer value and potential of the new
product concept, as well as to find new means to produce software.
Tyrväinen et al.~\cite{tyrvainen_metrics_2015} propose that measuring the cycle time from 
development to analysis of customer acceptance of the feature enables faster learning of 
market needs. In addition, receiving fast feedback from users makes changing the software 
easier for the programmers who have not yet forgotten the code.
Relevant research questions regarding internal software startups can be formulated as 
follows:
\begin{itemize}
	\item RQ1: How can Lean startup be adopted and adapted for software 
	product innovation in large software companies?
	\item RQ2: What are the challenges and enablers of Lean startup in large software 
	companies?
	\item RQ3: How should internal software startups be managed / lead?
	\item RQ4: What metrics can be used to evaluate software product innovation 
		in internal startups?
	\item RQ5: To what extent do internal startups have a competitive advantage 
	compared to independent startups (through shared resources, etc.)?

\end{itemize}
\medskip
Lean startup approach gains more interest from scholars and academics as a new way 
to foster innovation since it helps to avoid building products that nobody 
wants~\cite{Eisenmann2012}. Some evidence shows that mature software companies and 
startups differ in applying Lean startup approach~\cite{jarvinen_agile_2014}; e.g.
mature firms start the cycle by collecting data from existing users and then
generating a hypothesis based on that data, whereas software startups generate ideas and
collect data from new users to validate the ideas. However, it seems that, to a
large extent, the approach can be used both in startups and established
enterprises. By answering RQ1-RQ3 we aim at defining structured guidelines on how to 
introduce Lean startup in large software companies, supporting practitioners, while 
answering RQ4-RQ5 would provide a motivation for this approach, allowing to compare 
effectiveness on a quantitative level.

Due to the complex nature of the research phenomenon and the intention to achieve
an in-depth understanding of it, we consider multiple case studies~\cite{yin_case_2003} 
as a suitable research approach. The case organizations can be selected based on the 
following criteria: (1) the organization develops software in-house, (2) a dedicated team 
is responsible from ideation to commercialization of a new software, and (3) the software 
falls out of the current main product line. The unit of analysis in this study would be a 
development team.

Very few studies have investigated how the Lean startup~\cite{ries_lean_2011} can 
leverage internal startups in large software companies to improve their 
competency and capabilities of product innovation. Initial steps have been taken and some 
of the results have been published to fill this observed gap (e.g.~\cite{Edison2015a, 
Edison2015}). Marijarvi et al.~\cite{marijarvi2016} report on Finnish large companies’ 
experience in developing new software through internal startups. They also discuss the 
lifecycle phases of innovation work in large companies. The authors argue that different 
types of internal organization may take place in each stage of new product development. 
For example, problem/solution fit can be done in an internal startup or company 
subsidiary.

\subsubsection{Lean Startup for project portfolio management and open 
innovation}\label{PPMandOI}
Building on the challenges proposed in Section~\ref{inernalstartup}, we
propose that Lean startup could also be applied within both (i) project 
portfolio management (PPM), to co-ordinate multiple startup initiatives within 
an organization, and (ii) open innovation, wherein internal startups involve 
multiple organizations, individuals, or even unknown participants. Both PPM and 
open innovation and their main challenges are briefly introduced below, 
followed by research questions that require investigation before 
Lean startup principles can be successfully applied in these new contexts.

Software engineering PPM describes the ongoing identification, selection, 
prioritization, and management of the complete set of an organization's software 
engineering projects, which share common resources in order to maximize returns 
to the organization and achieve strategic business 
objectives~\cite{Meskendahl2010, cooper1999, Blichfeldt2008, 
turner2014handbook}. Open innovation is defined as the use of ``purposive inflows and 
outflows of knowledge to accelerate internal innovation and to expand the markets for 
external use of innovation, respectively''~\cite{chesbrough2003b}. Popular examples 
of open innovation include open source software development, crowd-sourcing, and inner 
source.

Effective PPM is critical to achieving business 
value~\cite{Hatzakis, Reyck2005}, improving cost and time savings, and 
eliminating redundancies~\cite{Kersten2003, LeFave2008}. Unfortunately, 
existing portfolio management practices, which are based on the effective completion of 
individual projects with only episodic portfolio level reviews~\cite{Reyck2005}, 
fail to manage either the dynamic nature of contemporary projects, or problems 
associated with portfolios comprising too many 
projects~\cite{Krebs2008,Reyck2005}. Indeed, many portfolios report an 
unwillingness to cancel projects that no longer contribute to the achievement 
of strategy~\cite{Reyck2005}. 

Open innovation (OI) presents numerous advantages for organizations, such as access to a 
requisite variety of experts, a prospective reduction in overall R\&D spending, reduced 
time-to-market, improved software development processes, and the integration of the firm 
into new and collaborative value networks~\cite{chesbrough2003b, Anbardan2014, 
Vanhaverbeke2006}. Nonetheless, adopting open innovation processes 
can be significantly challenging. For example, adopters often lack internal commitment, 
in addition to challenges associated with aligning innovation strategies to extend beyond 
the boundaries of the firm. Moreover, there are concerns regarding intellectual property 
and managing unknown contributors/contributions, as well as managing the higher costs and 
risks associated with managing both internal and external 
innovations~\cite{VandeVrande2009, West2006, Chesbrough2006}. The role of Lean startup 
principles in addressing these challenges in both PPM and OI is worthy of further 
research:

\begin{itemize}
	\item RQ1: How can Lean start-up be implemented within a portfolio management or 
	open innovation context?
	\item RQ2: How can Lean startup initiatives drive or accelerate open innovation?
	\item RQ3: What Lean startup concepts could be adapted to facilitate open 
	innovation processes in an organization?
	\item RQ4: How can one ensure Lean startup initiatives conducted across multiple 
	projects or organizations align with strategy?
	\item RQ5: How do you reconcile potential conflicts between portfolio / open 
	innovation processes and Lean startup processes?
	\item RQ6: How do you achieve consensus in defining the minimum viable product (MVP) 
	in networks comprised of multiple autonomous (and sometime anonymous) agents?
\end{itemize}
\medskip

The successful application of Lean startup principles (RQ1-RQ3) has the potential to 
reduce the costs arising from the poor implementation of PPM and OI practices and 
increase the value achieved from these initiatives. However, because such approaches are 
often practice led, it is necessary for academic research to develop effective theory to 
underpin practice and provide empirical data to support, or refute claims of 
effectiveness (RQ4-RQ6). Rich human interactions are at the heart of software engineering 
PPM and open innovation. Accordingly, phenomena in these domains can be examined using 
interpretive, qualitative methods such as semi-structured interviews, case studies and 
ethnography.

While the principles of lean have been applied to PPM 
(e.g.~\cite{hu_multi-objective_2008, cusumano_thinking_1998}, there is little research 
looking at the application of Lean startup principles to PPM. Similarly, while there is 
interest in the application of Lean startup principles in open innovation contexts, to 
date, such applications have predominantly been driven by practice. 

\subsection{Software Startup Ecosystems and Innovation Hubs}\label{sec:cluster5}
Successful software startups do not live in isolation. Normally, they are 
inserted in a rich environment that includes a number of relevant players, such 
as entrepreneurs, developers, investors, scientists, as well as business and 
intellectual property consultants. To support these players, a number  of 
support programs from the private and public sectors are required to provide 
funding, incubation, acceleration, training, networking, and consulting. All 
these elements combine into what scholars and practitioners have called Startup 
Ecosystems~\cite{kon2015}. In our software startups research agenda, we focus
on Software Startup Ecosystems (SSE) and the elements that are relevant for
startups that have software as a key part of their products or services.

By studying how SSEs are created, their main 
characteristics, and how they can evolve, one can better understand the environments that 
favour, or not, the birth and development of successful 
software startups. Research in this field can provide, to the relevant 
stakeholders, the concrete actions (e.g., public policies, 
private activities) that will establish a fruitful and vibrant environment 
for the execution of high-growth innovative projects within nascent software 
companies. The main research questions that need to be answered are the 
following:
\begin{itemize}
	\item RQ1: What are the key elements of a fruitful SSE?
	\item RQ2: Are there different types of SSEs, e.g. differentiated by size, technology 
	sectors, country economy or other factors? 
	\item RQ3: How do SSEs evolve over time?
	\item RQ4: How can one measure the output and qualities of an SSE? 
\end{itemize}

By answering RQ1, researchers will provide a better understanding of the way 
how SSEs and innovation hubs work, instrumenting key stakeholders in taking 
actions to improve their ecosystems. By identifying what factors promote or hinder the 
development of successful startups within a certain SSE, policy makers will get support 
in decision making (RQ2).  Entrepreneurs will also be able to better understand 
what are the environmental factors and forces that can help or hinder the success of
their enterprises.

Researchers from Brazil, Israel, and the USA have developed a methodology
to map a specific software startup ecosystem; this methodology has been
applied to Israel~\cite{kon2015}, S\~{a}o Paulo~\cite{Fonseca2015} and New 
York~\cite{Cukier2016}. Currently,
with the help of dozens of experts worldwide, they are developing a
maturity model for SSEs~\cite{kon2015,Cukier2015}, addressing RQ3 and RQ4. This maturity 
model needs further research and validation before it can be applied in real scenarios to 
help practitioners and policy makers.

The Global Startup Ecosystem Ranking~\cite{Herrmann2015} is crafted by a group 
of experts that have been proposing metrics to evaluate regional ecosystems 
around the world and compare them according to multiple criteria. Frenkel and 
Maital~\cite{Frenkel2014} have developed a methodology to map national 
innovation ecosystems and use this map to propose policies to promote 
improvement. 
Jayshree has studied the influence of environmental factors on entrepreneurial 
success~\cite{Jayshree2012}.
Finally, Sternberg~\cite{Sternberg2013} researched the role of regional 
government support programs and the regional environment as success factors 
for startups. 

\subsection{Theory and Methodologies for Software Startup 
Research}\label{sec:cluster6}
The tracks in this cluster direct their research towards identifying 
means to better study and understand software startups.

\subsubsection{Overview of the possible theoretical lenses for studying 
software startups}

Theories are important in any scientific field, as they form the foundation to 
understand a contemporary phenomenon better. Theories provide answers to the 
``why'' questions, and are therefore useful for explaining why certain events occur 
while others do not. Software startup research does not operate in a vacuum, 
but rather can borrow theories from both the software engineering and 
information systems fields, business and management literature, as well as from 
the fields of organizational and social sciences.

We have identified a few potential theories that can be meaningfully applied in 
the context of software startup companies. The proposed theories are the 
hunter-gatherer model~\cite{steinert_finding_2012}, Cynefin 
model~\cite{snowden_leaders_2007}, Effectuation theory~\cite{Sarasvathy2001} 
and Boundary Spanning theory~\cite{williams_competent_2002}. These theories are 
briefly outlined in this section.

Although 90\% of human history was occupied by hunters and gatherers, who forged for
wild plants and killed wild animal to survive, only 
recently was the hunter-gatherer model re-discovered by Steinert and 
Leifer~\cite{steinert_finding_2012} to explain how designers pursue 
their endeavours in search of the best design outcome. The model shows the 
changes in the design process, as well as subsequently in the design outcome. 
The model portrays a distinction between a hunter who aims to find an 
innovative idea, and a gatherer who aims to implement the idea. Both are needed 
to achieve concrete results. While hunting the idea through ambiguous 
spaces has a change-driven, analytical, and qualitative nature; gathering the idea 
across predetermined paths has a plan-oriented, manageable, and quantitative 
nature. The model has recently been applied in software startup research  to 
explain startups' evolutionary paths~\cite{NguyenDuc2015}.

Complexity theory has been used as a frame of reference, by analysing its 
implications on software design and development (e.g. 
Pelrine~\cite{pelrine_understanding_2011}, Rikkilä et 
al.~\cite{rikkila_implications_2012}). Software projects can be characterized as 
endeavours wherein a dynamic network of customers, software designers, 
developers, 3rd party partners, and external stakeholders interact and can 
be seen as a Complex Adaptive System (CAS).  To reason about decision-making in 
different situations, Snowden et al.~\cite{snowden_leaders_2007} proposed a 
sense-making framework for such systems. The model has five sub-domains and 
divides the world in two parts - ordered and unordered main domains. The 
ordered domain is the one in which cause-effect (CE) relationships are known (the 
Known domain), or at least knowable after analysis (the Complicate domain). In 
contrast, the unordered domain includes a complexity situation, wherein the CE 
relationship can only be perceived in retrospect, but not in advance (the 
Complex domain), and a chaotic situation, wherein behaviours are completely 
random, lacking any expected consequence when acted upon. Depending on the 
problem domain, suitable approaches include categorizing, analysing, probing 
or acting~\cite{snowden_leaders_2007}. The Cynefin model provides a framework that can be 
used to analyse the decisions made by software startuppers in developing their products. 
Often they find themselves in the unordered domain, attempting to make sense out of the 
current situation and navigate to the ordered domain.

Effectuation theory is a simple model, rooted in entrepreneurship, of decision-making 
under uncertainty. The effectual thinking is in the opposite of causal reasoning which 
starts from desired ends to necessary means (top-down). Experienced entrepreneurs reason 
from means to ends (bottom-up), trying to work out meanings and goals based on the 
resources they have at hand. The theory is embodied by 
five principles: the bird-in-hand principle, the affordable loss principle, the 
crazy quilt principle, the lemonade principle, and the pilot-in-the-plane 
principle~\cite{Sarasvathy2001}. The effectuation theory can help to 
make better sense of entrepreneurs' 
decision-making process in the evolution of software 
startups, such as problem validation, value proposition definition, design of 
MVPs, and pivoting processes. Good practices could be discovered using the 
effectuation theory as a theoretical lens.

Startups operate in a dynamic environment and face expectations and influences 
from many directions. In order to survive, they need to effectively collaborate  
within their team, but also outside it. Boundary spanning is a concept that deals 
with the structures of organizations that are transitioning from a rigid hierarchical 
structure 
towards a network-based expert organization, which gives rise to informal 
boundaries rather than structural ones~\cite{williams_competent_2002}. Boundary 
spanners are those people and entities who bridge these boundaries and 
opportunities. In the software engineering context, boundary spanning has been 
studied in the context of global software 
development~\cite{johri_boundary_2008}. Startuppers can be seen as boundary spanners 
when they need to 
bridge between various stakeholders. While 
boundaries are always unavoidable, but also necessary and useful, knowledge is 
required on how they can be crossed, rearranged, or even dissolved when considered 
harmful~\cite{wenger_communities_1998}. Startuppers should see boundaries 
as tools that facilitate and support making sense out of the environment. Boundary 
spanning helps in discovering how to overcome the challenges of distributed global 
work, where motivations, work styles, and knowledge domains vary across 
boundaries. Startuppers can become knowledge brokers, transferring and sharing their 
knowledge. 

There are other theoretical lenses that can be used to study software 
startups. Startups deal with innovative services and products, often for new or 
emerging markets. Birkinshaw et al.~\cite{birkinshaw_management_2008} analyse 
the innovation theories presented and propose a framework for management 
innovation process. This could be applied to the startup innovation process context 
to explore how product development moves from problem-driven search through 
trial and error to a finished prototype. The analysis can be complemented with 
Van de Ven and Poole’s~\cite{ven_explaining_1995} four views into 
organizational changes, in which they present alternate processes for 
organizations to transform.

Theorizing software startups is important, since there is a current lack of 
understanding of the dynamics in startups. Theoretical advancements need to be 
achieved so that researchers can make better sense out the diverse contexts, situations, 
and places where startuppers strive for success. 

\subsubsection{Defining the Lean Startup concept and evaluating practice}
Many positive drivers underpin the Lean Startup movement. The 
literature is abound with claims of reduced 
risk~\cite{Eisenmann2012,ries_lean_2011}, the benefits of evidenced-based 
trials~\cite{Blank2013, ries_lean_2011}, and shorter time-to-
market~\cite{ries_lean_2011}. 
We certainly know that these benefits are needed, given the challenges 
experienced by early stage software startups~\cite{Giardino2014, giardino_key_2015} and 
the percentage that fail~\cite{ries_lean_2011}. 
Indeed, many software start-ups fail~\cite{mullins_getting_2009, Crowne2002} because 
they waste too much time and money building the wrong product before realising too late 
what the right product should have been~\cite{nobel_teaching_2011, Bosch2013}. These 
challenges coupled with high uncertainty make the Lean Startup Methodology 
attractive to software startups as it supposedly offers an integrated approach 
to creating products and services that fit the market~\cite{harb_evaluating_2015}. This 
research builds on previous research conducted by Dennehy, Kasraian, 
O’Raghallaigh, and Conboy~\cite{dennehy_product_2016}, which identified a significant 
absence of frameworks that assisted startups to efficiently and effectively progress 
their Minimum Viable Products (MVP) to a Product Market Fit (PMV).  The theoretical 
advancement of the lean concept in contemporary software engineering and software 
development literature has been arrested, mainly because the academic research community 
has followed ``fads and fancies'' which characterize academic research. The implications 
for the arrested theoretical development of lean concept, listed next, are the motivation 
for this research.

As is often the case with new and emerging phenomena, Lean Startup practice has led
research, with the creation, promotion, and dissemination of these methods 
almost completely due to the efforts of practitioners and consultants. Now, Lean 
Startup research is beginning to gain momentum, as is evident from the 
increasing number of dedicated journal special issues, conferences, conference 
tracks, and workshops. While there are merits to adopting such a 
practice-oriented focus, little if any research effort has focused on the 
conceptual development of Lean Startup and its underlying components. As 
practice has lead research, the definition of Lean Startup has emerged through 
how it is used in practice. As a result, Lean Startup adoption is often defined 
by how the practices are adhered to, rather than the value gleaned from their 
use, adaptation, or, in some cases, abandonment. We see this in many other methods 
such as in agile, where many define ``being agile'' as how many Scrum or XP 
practices are used, rather than the value obtained by their 
use~\cite{Conboy2009}. As a result, the current body of software startup 
knowledge suffers from a number of limitations, including:

(1) Lack of clarity: While there is broad agreement in principle regarding what 
constitutes key concepts such as MVP, assumptions regarding the specific definitions, 
interpretations, use, and evaluations are often unclear in many existing Lean Startup 
studies. This makes critical appraisal, evidence-based evaluation, and comparison 
across 	studies extremely difficult.

(2) Lack of cohesion and cumulative tradition: A good concept or theory should 
cumulatively build on existing research. Very little academic research has examined Lean 
Startup using concepts that have more mature and substantive bodies of research with 
theories, frameworks and other lenses that have been thoroughly tested over time. The 
lean concept has been applied in manufacturing since WW1, and yet in Lean Startup 
research we see very myopic and limited use of the broad lean frameworks available. Other 
concepts that influence Lean Startup include agility, flow, and innovation.

(3) Limited applicability: Adherence-based measures of Lean Startup inhibit the ability 
to apply Lean Startup in domains other than that originally intended. Research now 
attempts to apply Lean Startup in other environments, such as large organizations and 
regulated environments, and so this will become a more prevalent issue as this 
trend continues. Therefore, questions relevant for this research track include:
\begin{itemize}
	\item RQ1: What are the core concepts that underpin Lean Startup?
	\item RQ2: What are the components of a higher abstract Lean Startup that 
	allows the concept to be applied and evaluated in a value-based manner?
	\item RQ3: What theories, frameworks, metrics, and other instruments from 
	these existing related bodies of knowledge can be applied to Lean Startup?
	\item RQ4: How can these be effectively applied to improve the use of Lean 
	Startup in practice, and the study and improvement of Lean Startup in research?
	\item RQ5: How can Lean Startup then be tailored to suit environments it 
	was not originally designed to support, e.g. large organizations, 
	regulated environments, or peer production?
	\item RQ6: Does Lean Startup enable or inhibit fundamental leaps in business 
	and software business ideas? For example, does MVP place an invisible 
	ceiling, wherein once you reach MVP  you subconsciously stop looking for the 
	truly significant innovation?
\end{itemize}
\medskip
As there is reciprocal relationship between practice and academia, where academic 
research is informed by practice and practice is informed by academic research, this 
research would impact on research and on practice. By answering RQ4-RQ6, this research 
track would provide practice with empirical evidence on the utility of lean practices in 
diverse environments, while also positioning the lean method at the core of academic 
research (RQ1-RQ3). As case study research is an empirical inquiry that ``investigates a 
contemporary phenomenon in depth and within its real-life context''~\cite{yin_case_2003}, 
it would be highly suited to addressing the theoretical limitations of lean and for 
answering the questions listed above. Specifically, the use of a multiple-case design 
would allow a cross-case pattern to develop more sophisticated descriptions and powerful 
explanations~\cite{miles_qualitative_1994} of the lean concept. 

The challenges of new product development are not confined to software 
startups. Therefore, software engineering teams working in distributed or regulated 
environments such as financial services and within multinational companies would provide 
rich insights to the advancement of the lean concept.

\subsubsection{Research collaboration strategies with software startups}
\label{sub:rescol}
Empirical research in the area of software engineering normally requires access to 
organizations and artefacts from companies 
developing software intensive products and services~\cite{WohlinClaes2012}. 
In the case of startups, such access is very limited, due to several challenges:
\begin{enumerate}
	\item startups have limited resources both in terms of person hours and 
	calendar time for anything but working on their MVP,
	\item startups want all investments to yield almost immediate results, 
	thus investments in long-term potential are not prioritized, and
	\item artefacts and actual products are often very sensitive, as the 
	startup is very vulnerable.
\end{enumerate}
These and other reasons limit
empirical research, as reflected in both academic knowledge 
about startups overall, but also in the superficial nature of what is available. 
For this reason, any initiative to seriously collect empirical data as well as
conduct research on core challenges facing startups has to originate with a 
strategy that overcomes these obstacles.
One possible strategy is to pool resources and access to startups, in essence 
sharing empirical data and coordinating research into startup software engineering. 
Coordination should be seen as equally central, as it enables researchers to limit 
the impact and costs as each study and project part can be focused and 
small, and several larger issues can be tackled through coordination. 
Concrete examples of joint activities include, but are not limited to:
\begin{enumerate}
	\item joint surveys at the superficial level (pooling resources to collect 
	many data points),
	\item complementary surveys and case studies where each partner does a 
	part only, but the results can be combined in analysis and synthesis,
	\item formulating a complementary research agenda with clear interfaces 
	and joint research questions, and
	\item pooling resources in relation to testing ``solutions'' emerging from 
	the collaboration.
\end{enumerate}

While this strategy opens the possibility to share the resource requirements 
among the studied startups, there are open questions regarding its implementation:
\begin{itemize}
	\item RQ1: To what extent is data from different startups and startup 
	ecosystems comparable? In other words, which techniques exist to 
	perform meta-analysis of the gathered heterogeneous data?
	\item RQ2: How can we efficiently transfer technology between researchers and 
	startups, and how can we measure the impact of transferred solutions?
\end{itemize}

We conjecture that the software startup context model discussed in 
Section~\ref{sec:context} would be an enabler for answering RQ1. Confounding 
variables~\cite{unterkalmsteiner_conceptual_2014} 
could then be easier identified, allowing for sample stratification and robust 
statistical analyses~\cite{kitchenham_robust_2016}. In particular, data collected from 
different researchers could be aggregated and increase the strength of the conclusions 
drawn from the analysis, i.e. enabling meta-analysis~\cite{hayes_research_1999}.

Answering RQ2 would allow us to actually support software startups on a broad basis with 
the knowledge gained from the research proposed in this agenda. While different 
approaches exist to transfer knowledge from academia to 
industry~\cite{gorschek_model_2006, wieringa_empirical_2014}, they are mostly targeted at 
mature companies that have the resources to collaborate with researchers over a longer 
period of time. We think that software startup ecosystems, discussed in 
Section~\ref{sec:cluster5}, can contribute to technology transfer if researchers are 
active in these structures and can create a win-win situation where both startups and 
researchers benefit.

\section{Discussion}\label{discussion}
In this section we give a brief overview of the research tracks in relation to 
other work in software engineering and their potential impact on the field. We conclude 
this section with a discussion on the study's limitations.

Software startup engineering research centers around the core knowledge 
base in Software Engineering~\cite{bourque_guide_2014}. This is illustrated by 
the research tracks proposed in Section~\ref{sec:cluster1} that encompass 
providing support for startup engineering activities. Noticing what is considered 
``good'' software engineering practice~\cite{bourque_guide_2014}, and the challenges that 
software startups encounter~\cite{giardino_key_2015, klotins_software_2016}, 
we see potential in directing research towards efficient and effective 
requirements for engineering practices in startups. Klotins et 
al.~\cite{klotins_software_2016} studied 88 experience reports from startups and 
identified lack of requirements validation, classification (to enable 
prioritization), and identification of requirements sources (to identify a 
relevant value proposition) as causes for engineering uncertainty, which maps 
to the early-stage startup challenges of technology uncertainty and delivering 
customer value, identified by Giardino et al.~\cite{giardino_key_2015}. 
Unlike large companies, software startups have unique time and resource constraints and 
thus cannot afford to develop features and services that will not be used or valued by 
the customers. We believe that lightweight practices to identify, and, most importantly, 
analyse requirements for their business value can help software startups in their 
decision process. Looking at the research tracks in Section~\ref{sec:cluster1}, several 
of them touch upon requirements engineering aspects. Prototypes can be used to 
communicate with customers to elicit requirements (Section~\ref{sec:prototype}), while 
product innovation assessment (Section~\ref{sec:innovation}) is relevant in the context 
of analysing the customers' perceived value of the offered solutions. Even optimizing the 
effort spent in requirements engineering and quality assurance, for example by using test 
cases as requirements~\cite{bjarnason_multi-case_2016}, involving product users for 
testing (Section~\ref{sec:testing}), addresses requirements engineering aspects.  

The focus on requirements in software startup engineering research directly relates to 
the research tracks presented in Section~\ref{sec:cluster2}, startup evolution models and 
patterns, as the cost of pivoting could be reduced by earlier and less ad-hoc analysis of 
requirements and value propositions of the envisioned products. The patterns emerging 
from the research on survival capabilities of software startups, proposed in 
Section~\ref{sub:survival}, could provide valuable heuristics leading to a lightweight 
analysis of product value propositions. The research on pivoting and survival 
capabilities is likely to affect software startup practitioners on a strategic level by 
providing them managerial decision support that draws from models rooted in software 
engineering practice. An example where such a cross-discipline approach has been very 
successful is value-based software engineering~\cite{boehm_value-based_2003}.  

The research tracks described in Section~\ref{sec:cluster3} were grouped under the name 
``cooperative and human aspects in software startups'', borrowed from the research area 
in software engineering that is interested in studying the impact of cognitive abilities, 
team composition, workload, informal communication, expertise identification and other 
human aspects on software construction~\cite{souza_guest_2009}. We conjecture that 
studying and understanding these aspects better has a large potential as software 
startups are driven by motivated individuals rather than a corporate agenda. Lessons from 
this research can both benefit startup practitioners, in particular in conjunction with 
the work on software startups ecosystems (Section~\ref{sec:cluster5}), and more mature 
companies, for example by applying models of competency needs that could emerge from the 
work presented in Section~\ref{sec:competency}.

The remaining research tracks described in 
Sections~\ref{sec:cluster4}~-~\ref{sec:cluster5} take a step back from what 
happens \emph{inside} a software startup. The research tracks in 
Section~\ref{sec:cluster4} propose to apply startup concepts in non-startup 
contexts. The idea of extracting a concept from one context and applying it in 
another has proven successful in other areas, such as in systematic literature 
reviews~\cite{mulrow_rationale_1994,kitchenham_evidence-based_2004} and open 
source principles~\cite{torkar_adopting_2011, 
hippel_innovation_2001,west_how_2003}. The premise of internal startups is that 
the positive traits of ``startups in the wild'' can be transferred to a 
corporate environment, fostering innovation and faster product development. The 
overall aim of the research tracks described in Section~\ref{sec:cluster4} is 
to evaluate whether the traits of startups can actually produce thriving 
environments within mature companies. In comparison, the research on startup 
ecosystems and innovation hubs (Section~\ref{sec:cluster5}) takes a broader and 
higher level view of software startup phenomenon. Neither independent startups 
nor mature companies adopting internal startup initiatives live in 
isolation. A better understanding of startup ecosystems and innovation hubs 
might thereby provide key insights into the factors that create a fruitful 
software startup environment.

Finally, the research tracks in Section~\ref{sec:cluster6} look at aspects 
relevant for implementing the research agenda described in this paper. In 
particular, theories that can be used to better understand the dynamics in and 
around software startups are of value when attempting to construct a more holistic 
understanding of software startups in their various contexts.
For the research on defining the Lean Startup concept, parallels to and lessons 
from similar endeavours around research on agile software 
development~\cite{dyba_empirical_2008} should be taken into consideration. In this paper, 
we followed a recommendation by Dybå and Dingsøyr to develop a research agenda on the 
phenomenon of interest~\cite{dyba_empirical_2008}. However, in order to implement this 
research agenda, we need to also answer the questions about how to enable 
efficient and effective research collaborations with software startups 
(Section~\ref{sub:rescol}).

\subsection{Limitations}
The research agenda presented in this paper was developed ``bottom-up'', i.e. the areas 
of interest were proposed and described by a sample of software startup researchers 
without any restriction on covering certain aspects of the software engineering body of 
knowledge but guided by their past, current and future work in the field. Often, these 
researchers have both a leg in academia and in the startup community, either as mentors, 
founders, or simply as part of the development team. This approach to develop a research 
agenda is not uncommon (see e.g.~\cite{broy_challenges_2006, chandra_software_2014, 
cleland-huang_software_2014}), but is threatened by a potential 
bias towards the preferences of individual researchers. This is why we invited a large 
number of our peers to contribute to the agenda. Even though the research tracks cover 
many software engineering aspects and beyond, the agenda is only a sample of the 
potentially relevant future research on software startups. This means that potentially 
interesting and relevant research topics, such as use of open source software, business 
model development, legal issues and intellectual property rights, are not discussed in 
this paper. However, we expect that the agenda will grow together with the research 
community as soon as the work on the proposed research tracks bears fruits, leading to 
new research questions.

\section{Outlook and Conclusions}
\label{outlook}
Software startups are an interesting and stimulating phenomenon in the modern 
economy and are of paramount importance for the societies of today. Despite of high 
failure rates, communities, cities and countries are investing on stimulating the 
creation of software startups. While these startups may not solve the unemployment 
problems of many countries they stimulate a new type of positive dynamism in societies 
encouraging people to collaborate and develop their personal skills in novel ways.
The emergence of the software startup research area reflects the fact 
that we need to better understand this phenomenon to learn valuable 
lessons and accumulate valid knowledge to benefit future 
entrepreneurial initiatives. The research agenda described in this paper is one 
of the first attempts to establish the software startup as a nascent, yet fast 
growing research area, and to depict its landscape by highlighting the 
interesting research topics and questions to explore.

It is worth emphasizing again that software engineering is only one of the 
multiple disciplines that are relevant and can inform software startup 
practice. Other disciplines include Economics, Entrepreneurship, Design, 
Finance, Sociology, and Psychology. Therefore, there is a need to 
collaborate with researchers from these disciplines in order to increase the potential 
of achieving relevant and useful research results that can benefit practice. 

Due to the emerging nature of the field, there is still much to be done to 
establish software startups as a research area. Relevant concepts need clear 
definitions, substantive theories need to be developed, and initial research 
findings need to be validated by future studies. Software startups are very 
diversified in terms of entrepreneurs' varying
approaches to their startup endeavours. Without the sound foundation mentioned above 
for this research area, there are risks of asking irrelevant research 
questions and not being able to attain rigorous results. 

Last but not least, this research agenda is not meant to be exhaustive, and we 
are aware that we may exclude some important Software Engineering topics relevant to 
software startups. The research agenda is open to additions of new tracks, topics, and 
research questions by other researchers interested in the research area. With  
contributions and commitments from researchers from different institutions 
and backgrounds, collectively we can establish software startup as a promising 
and significant research area that attracts more exciting discovery and 
contribution. We welcome those interested in joining the Software Startup Research 
Network in fostering the collaboration between researchers and taking the research agenda 
further.

\bibliography{references}

\begin{thebibliography}{100}

\bibitem{isoiec25000}
Software engineering -- software product quality requirements and evaluation
  ({SQuaRE}) -- guide to {SQuaRE} - {ISO/IEC} 25000:2005.
\newblock Technical report, International Organization for Standardization,
  2005.

\bibitem{_iso/iec/ieee_2011}
{ISO}/{IEC}/{IEEE} {Systems} and software engineering – {Architecture}
  description.
\newblock {\em ISO/IEC/IEEE 42010:2011(E) (Revision of ISO/IEC 42010:2007 and
  IEEE Std 1471-2000)}, pages 1--46, 2011.

\bibitem{sba2014}
Frequently asked questions about small business.
\newblock Technical report, {U.S.} Small Business Administration, 2014.

\bibitem{Anbardan2014}
Y.~Z. Anbardan and M.~Raeyat.
\newblock Open {Innovation}: {Creating} {Value} {Through} {Co}-{Creation}.
\newblock In {\em Proceedings 7th {World} {Conference} on {Mass}
  {Customization}, {Personalization}, and {Co}-{Creation} ({MCPC} 2014)}, pages
  437--447, Aalborg, Denmark, 2014. Springer.

\bibitem{618169}
R.~Balachandra and J.~Friar.
\newblock Factors for success in {R\&D} projects and new product innovation: a
  contextual framework.
\newblock {\em IEEE Transactions on Engineering Management}, 44(3):276 --287,
  1997.

\bibitem{Bart1988}
C.~K. Bart.
\newblock {New venture units: use them wisely to manage innovation}.
\newblock {\em Sloan Management Review}, 29(4):35--43, 1988.

\bibitem{beaudouin-lafon_prototyping_2002}
M.~Beaudouin-Lafon and W.~E. Mackay.
\newblock Prototyping development and tools.
\newblock {\em Handbook of Human-Computer Interaction}, pages 1006--1031, 2002.

\bibitem{birkinshaw_management_2008}
J.~Birkinshaw, G.~Hamel, and M.~J. Mol.
\newblock Management innovation.
\newblock {\em Academy of management Review}, 33(4):825--845, 2008.

\bibitem{bjarnason_multi-case_2016}
E.~Bjarnason, M.~Unterkalmsteiner, E.~Engström, and M.~Borg.
\newblock A {Multi}-{Case} {Study} of {Agile} {Requirements} {Engineering} and
  {Using} {Test} {Cases} as {Requirements}.
\newblock {\em Information and Software Technology}, 77:61--79, 2016.

\bibitem{Blank2005}
S.~Blank.
\newblock {\em {The Four Steps to the Epiphany: Successful Strategies for
  Products that Win}}.
\newblock Cafepress.com, 2005.

\bibitem{Blank2013}
S.~Blank.
\newblock {Why the Lean Start-Up Changes Everything}.
\newblock {\em Harvard Business Review}, 91(5), 2013.

\bibitem{Blank2012}
S.~Blank and B.~Dorf.
\newblock {\em {The Startup Owner's Manual: The Step-By-Step Guide for Building
  a Great Company}}.
\newblock K {\&} S Ranch, 2012.

\bibitem{Blichfeldt2008}
B.~S. Blichfeldt and P.~Eskerod.
\newblock {Project portfolio management - There's more to it than what
  management enacts}.
\newblock {\em International Journal of Project Management}, 26(4):357--365,
  2008.

\bibitem{Block1985}
Z.~Block and I.~C. MacMillan.
\newblock {Milestones for successful venture planning}.
\newblock {\em Harvard Business Review}, 63(5):184--196, 1985.

\bibitem{boehm_value-based_2003}
B.~Boehm.
\newblock Value-based {Software} {Engineering}.
\newblock {\em SIGSOFT Softw. Eng. Notes}, 28(2):1--12, 2003.

\bibitem{Bosch2013}
J.~Bosch, H.~H. Olsson, J.~Björk, and J.~Ljungblad.
\newblock The {Early} {Stage} {Software} {Startup} {Development} {Model}: {A}
  {Framework} for {Operationalizing} {Lean} {Principles} in {Software}
  {Startups}.
\newblock In {\em Proceedings 4th {International} {Conference} on {Lean}
  {Enterprise} {Software} and {Systems} ({LESS})}, pages 1--15, Galway,
  Ireland, 2013. Springer.

\bibitem{bourque_guide_2014}
P.~Bourque and R.~E. Fairley, editors.
\newblock {\em Guide to the {Software} {Engineering} {Body} of {Knowledge}}.
\newblock IEEE, 3rd edition, 2014.

\bibitem{brem_integration_2009}
A.~Brem and K.-I. Voigt.
\newblock Integration of market pull and technology push in the corporate front
  end and innovation management—{Insights} from the {German} software
  industry.
\newblock {\em Technovation}, 29(5):351--367, 2009.

\bibitem{brown_change_2009}
T.~Brown.
\newblock {\em Change by {Design}: {How} {Design} {Thinking} {Transforms}
  {Organizations} and {Inspires} {Innovation}}.
\newblock HarperBusiness, New York, 2009.

\bibitem{broy_challenges_2006}
M.~Broy.
\newblock Challenges in {Automotive} {Software} {Engineering}.
\newblock In {\em Proceedings 28th {International} {Conference} on {Software}
  {Engineering} ({ICSE})}, pages 33--42, Shanghai, China, 2006. ACM.

\bibitem{brush_properties_2008}
C.~G. Brush, T.~S. Manolova, and L.~F. Edelman.
\newblock Properties of emerging organizations: {An} empirical test.
\newblock {\em Journal of Business Venturing}, 23(5):547--566, 2008.

\bibitem{Carmel1994}
E.~Carmel.
\newblock {Time-to-completion in software package startups}.
\newblock In {\em Proceedings 27th Hawaii International Conference on System
  Sciences ({HICSS})}, pages 498--507. IEEE, 1994.

\bibitem{chandra_software_2014}
S.~Chandra, V.~S. Sinha, S.~Sinha, and K.~Ratakonda.
\newblock Software {Services}: {A} {Research} {Roadmap}.
\newblock In {\em Future of {Software} {Engineering} ({FOSE})}, pages 40--54,
  Hyderabad, India, 2014. ACM.

\bibitem{Chen2009}
C.-J. Chen.
\newblock {Technology commercialization, incubator and venture capital, and new
  venture performance}.
\newblock {\em Journal of Business Research}, 62(1):93--103, 2009.

\bibitem{Chesbrough2006}
H.~Chesbrough and A.~Crowther.
\newblock {Beyond high tech: early adopters of open innovation in other
  industries}.
\newblock {\em R{\&}d Management}, 36(3):229--236, 2006.

\bibitem{chesbrough2003b}
H.~W. Chesbrough.
\newblock {\em Open innovation: The new imperative for creating and profiting
  from technology}.
\newblock Harvard Business Press, 2006.

\bibitem{clarke_situational_2012}
P.~Clarke and R.~V. O'Connor.
\newblock The situational factors that affect the software development process:
  {Towards} a comprehensive reference framework.
\newblock {\em Information and Software Technology}, 54(5):433--447, 2012.

\bibitem{clarysse_explaining_2011}
B.~Clarysse, J.~Bruneel, and M.~Wright.
\newblock Explaining growth paths of young technology-based firms: structuring
  resource portfolios in different competitive environments.
\newblock {\em Strategic Entrepreneurship Journal}, 5(2):137--157, 2011.

\bibitem{cleland-huang_software_2014}
J.~Cleland-Huang, O.~C.~Z. Gotel, J.~Huffman~Hayes, P.~Mäder, and A.~Zisman.
\newblock Software {Traceability}: {Trends} and {Future} {Directions}.
\newblock In {\em Proceedings {Future} of {Software} {Engineering}}, pages
  55--69, Hyderabad, India, 2014. ACM.

\bibitem{cocco_simulating_2011}
L.~Cocco, K.~Mannaro, G.~Concas, and M.~Marchesi.
\newblock Simulating {Kanban} and {Scrum} vs. {Waterfall} with {System}
  {Dynamics}.
\newblock In {\em Proceedings 12th {Internation} {XP} {Conference} ({XP})},
  pages 117--131, Madrid, Spain, 2011. Springer.

\bibitem{Coleman2008}
G.~Coleman and R.~V. O'Connor.
\newblock {An investigation into software development process formation in
  software start-ups}.
\newblock {\em Journal of Enterprise Information Management}, 21(6):633--648,
  2008.

\bibitem{COLEMAN2013}
S.~Coleman, C.~Cotei, and J.~Farhat.
\newblock {A Resource-based view of new firm survival: new perspectives on the
  role of industry and exit route}.
\newblock {\em Journal of Developmental Entrepreneurship}, 18(01):1--25, 2013.

\bibitem{Conboy2009}
K.~Conboy.
\newblock {Agility from first principles: Reconstructing the concept of agility
  in information systems development}.
\newblock {\em Information Systems Research}, 20(3):329--354, 2009.

\bibitem{concas_simulation_2013}
G.~Concas, M.~I. Lunesu, M.~Marchesi, and H.~Zhang.
\newblock Simulation of software maintenance process, with and without a
  work-in-process limit.
\newblock {\em Journal of Software: Evolution and Process}, 25(12):1225--1248,
  2013.

\bibitem{cooper1999}
R.~G. Cooper.
\newblock From experience - the invisible success factors in product
  innovation.
\newblock {\em Journal of Product Innovation Management}, 16(2):115--133, 1999.

\bibitem{Cooper2011}
R.~G. Cooper.
\newblock {Perspective: The innovation dilemma: How to innovate when the market
  is mature}.
\newblock {\em Journal of Product Innovation Management}, 28(SUPPL. 1):2--27,
  2011.

\bibitem{croll_lean_2013}
A.~Croll and B.~Yoskovitz.
\newblock {\em Lean {Analytics}: {Use} {Data} to {Build} a {Better} {Startup}
  {Faster}}.
\newblock O'Reilly Media, Sebastopol, USA, 1st edition, 2013.

\bibitem{Crossan_Apaydin_2009}
M.~M. Crossan and M.~Apaydin.
\newblock A multi-dimensional framework of organizational innovation: A
  systematic review of the literature.
\newblock {\em Journal of Management Studies}, 47(6):1154--1191, 2009.

\bibitem{Crowne2002}
M.~Crowne.
\newblock Why software product startups fail and what to do about it.
\newblock In {\em International Engineering Management Conference (IEMC)},
  pages 338--343, Cambridge, UK, 2002. IEEE.

\bibitem{Cukier2015}
D.~Cukier, F.~Kon, and N.~Krueger.
\newblock {Designing a Maturity Model for Software Startup Ecosystems}.
\newblock In {\em Proceedings 1st International Workshop on Software Startups},
  pages 600--606, Bolzano, Italy, 2015. Springer.

\bibitem{Cukier2016}
D.~Cukier, F.~Kon, and L.~S. Thomas.
\newblock {Software Startup Ecosystems Evolution: The New York City Case
  Study}.
\newblock In {\em Proceedings 2nd International Workshop on Software Startups},
  Trondheim, Norway, 2016. IEEE.

\bibitem{cusumano_thinking_1998}
M.~A. Cusumano and K.~Nobeoka.
\newblock {\em Thinking {Beyond} {Lean}: {How} {Multi}-project {Management} is
  {Transforming} {Product} {Development} at {Toyota} and {Other} {Companies}}.
\newblock Simon and Schuster, 1998.

\bibitem{dennehy_product_2016}
D.~Dennehy, L.~Kasraian, O.~O’Raghallaigh, and K.~Conboy.
\newblock Product {Market} {Fit} {Frameworks} for {Lean} {Product}
  {Development}.
\newblock In {\em Proceedings {R}\&{D} {Management} {Conference} 2016 “{From}
  {Science} to {Society}: {Innovation} and {Value} {Creation}”}, Cambridge,
  UK, 2016.

\bibitem{Dess2003}
G.~G. Dess, R.~D. Ireland, S.~A. Zahra, S.~W. Floyd, J.~J. Janney, and P.~J.
  Lane.
\newblock {Emerging Issues in Corporate Entrepreneurship}.
\newblock {\em Journal of Management}, 29(3):351--378, 2003.

\bibitem{Dingsoyr2013}
T.~Dings{\o}yr and Y.~Lindsj{\o}rn.
\newblock {Team Performance in Agile Development Teams: Findings from 18 Focus
  Groups}.
\newblock In {\em Proceedings 14th International Conference on Agile Software
  Development}, pages 46--60, Vienna, Austria, 2013.

\bibitem{dovan_reliability_1998}
K.~Dovan.
\newblock Reliability in content analysis: {Some} common misconceptions and
  recommendations.
\newblock {\em Human Communication Research}, 30(3):411--433, 1998.

\bibitem{dyba_a_what_2012}
T.~Dyb{\aa}, D.~I. Sj{\o}berg, and D.~S. Cruzes.
\newblock What {Works} for {Whom}, {Where}, {When}, and {Why}?: {On} the {Role}
  of {Context} in {Empirical} {Software} {Engineering}.
\newblock In {\em Proceedings {International} {Symposium} on {Empirical}
  {Software} {Engineering} and {Measurement} ({ESEM})}, pages 19--28, Lund,
  Sweden, 2012. ACM.

\bibitem{dyba_empirical_2008}
T.~Dybå and T.~Dingsøyr.
\newblock Empirical studies of agile software development: {A} systematic
  review.
\newblock {\em Information and Software Technology}, 50(9–10):833--859, 2008.

\bibitem{Edison2015}
H.~Edison.
\newblock {A Conceptual Framework of Lean Startup Enabled Internal Corporate
  Venture}.
\newblock In {\em Proceedings 1st International Workshop on Software Startups},
  pages 607--613, Bolzano-Bozen, Italy, 2015. Springer.

\bibitem{edison_towards_2013}
H.~Edison, N.~bin Ali, and R.~Torkar.
\newblock Towards innovation measurement in the software industry.
\newblock {\em Journal of Systems and Software}, 86(5):1390--1407, 2013.

\bibitem{edison_towards_2015}
H.~Edison, D.~Khanna, S.~S. Bajwa, V.~Brancaleoni, and L.~Bellettati.
\newblock Towards a {Software} {Tool} {Portal} to {Support} {Startup}
  {Process}.
\newblock In {\em Proceedings 1st International Workshop on Software Startups},
  pages 577--583, Bolzano, Italy, 2015. Springer.

\bibitem{Edison2015a}
H.~Edison, X.~Wang, and P.~Abrahamsson.
\newblock {Lean startup}.
\newblock In {\em Scientific Workshop Proceedings of the XP conference}, pages
  1--7, Helsinki, Finland, 2015. ACM.

\bibitem{Efeoglu2013}
A.~Efeoğlu, C.~Møller, and M.~Sérié.
\newblock Solution {Prototyping} with {Design} {Thinking} – {Social} {Media}
  for {SAP} {Store}: {A} {Case} {Study}.
\newblock In {\em Proceedings {European} {Design} {Science} {Symposium}
  ({EDSS})}, pages 99--110, Dublin, Ireland, 2013. Springer.

\bibitem{eisenhardt_dynamic_2000}
K.~M. Eisenhardt and J.~A. Martin.
\newblock Dynamic capabilities: what are they?
\newblock {\em Strategic Management Journal}, 21(10-11):1105--1121, 2000.

\bibitem{Eisenmann2012}
T.~R. Eisenmann, E.~Ries, and S.~Dillard.
\newblock {Hypothesis-Driven Entrepreneurship: The Lean Startup}.
\newblock {\em Harvard Business School}, 2012.

\bibitem{elo_qualitative_2008}
S.~Elo and H.~Kyngäs.
\newblock The qualitative content analysis process.
\newblock {\em Journal of Advanced Nursing}, 62(1):107--115, 2008.

\bibitem{Ensley2006}
M.~D. Ensley, K.~M. Hmieleski, and C.~L. Pearce.
\newblock {The importance of vertical and shared leadership within new venture
  top management teams: Implications for the performance of startups}.
\newblock {\em The Leadership Quarterly}, 17(3):217--231, 2006.

\bibitem{eversheim_innovation_2008}
W.~Eversheim.
\newblock {\em Innovation {Management} for {Technical} {Products}: {Systematic}
  and {Integrated} {Product} {Development} and {Production} {Planning}}.
\newblock Springer Science \& Business Media, 2008.

\bibitem{Feng2012929}
T.~Feng, L.~Sun, C.~Zhu, and A.~S. Sohal.
\newblock Customer orientation for decreasing time-to-market of new products
  {IT} implementation as a complementary asset.
\newblock {\em Industrial Marketing Management}, 41(6):929 -- 939, 2012.

\bibitem{fernandez_2015}
C.~Fern{\'a}ndez-S{\'a}nchez, J.~Garbajosa, and A.~Yag{\"u}e.
\newblock A framework to aid in decision making for technical debt management.
\newblock In {\em Proceedings 7th {International} {Workshop} on {Managing}
  {Technical} {Debt} ({MTD})}, pages 69--76, Bremen, Germany, 2015. IEEE.

\bibitem{Fonseca2015}
M.~C. Fonseca.
\newblock {O ecossistema de startups de software da cidade de S{\~{a}}o Paulo}.
\newblock Master's thesis, University of S{\~{a}}o Paulo, 2016.

\bibitem{Francis2000}
D.~Francis and W.~Sandberg.
\newblock {Friendship within entrepreneurial teams and its association with
  team and venture performance}.
\newblock {\em Entrepreneurship: Theory and Practice}, 25(2):5--21, 2000.

\bibitem{Frenkel2014}
A.~Frenkel and S.~Maital.
\newblock {\em {Mapping National Innovation Ecosystems: Foundations for Policy
  Consensus}}.
\newblock Edward Elgar Publishing, London, UK, 2014.

\bibitem{fuller_user_2013}
J.~F{\"u}ller, R.~Schroll, and E.~von Hippel.
\newblock User generated brands and their contribution to the diffusion of user
  innovations.
\newblock {\em Research Policy}, 42(6--7):1197--1209, 2013.

\bibitem{gans_product_2003}
J.~S. Gans and S.~Stern.
\newblock The product market and the market for “ideas”: commercialization
  strategies for technology entrepreneurs.
\newblock {\em Research Policy}, 32(2):333--350, 2003.

\bibitem{Garvin2006}
D.~A. Garvin and L.~C. Levesque.
\newblock {Meeting the challenge of corporate entrepreneurship.}
\newblock {\em Harvard business review}, 84(10):102--12, 150, 2006.

\bibitem{giardino_key_2015}
C.~Giardino, S.~S. Bajwa, X.~Wang, and P.~Abrahamsson.
\newblock Key {Challenges} in {Early}-{Stage} {Software} {Startups}.
\newblock In {\em Proceedings 16th {International} {XP} {Conference} ({XP})},
  pages 52--63, Helsinki, Finland, 2015. Springer.

\bibitem{giardino_software_2016}
C.~Giardino, N.~Paternoster, M.~Unterkalmsteiner, T.~Gorschek, and
  P.~Abrahamsson.
\newblock Software {Development} in {Startup} {Companies}: {The} {Greenfield}
  {Startup} {Model}.
\newblock {\em Transactions on Software Engineering}, 42(6):585--604, 2016.

\bibitem{Giardino2014b}
C.~Giardino, M.~Unterkalmsteiner, N.~Paternoster, T.~Gorschek, and
  P.~Abrahamsson.
\newblock What do we know about software development in startups?
\newblock {\em IEEE Software}, 31(5):28--32, 2014.

\bibitem{Giardino2014}
C.~Giardino, X.~Wang, and P.~Abrahamsson.
\newblock {Why Early-Stage Software Startups Fail: A Behavioral Framework}.
\newblock In {\em Proceedings 5th International Conference on Software Business
  (ICSOB)}, pages 27--41, Paphos, Cyprus, 2014. Springer.

\bibitem{giones_strategic_2015}
F.~Giones and F.~Miralles.
\newblock Strategic {Signaling} in {Dynamic} {Technology} {Markets}: {Lessons}
  {From} {Three} {IT} {Startups} in {Spain}.
\newblock {\em Global Business and Organizational Excellence}, 34(6):42--50,
  2015.

\bibitem{gorschek_model_2006}
T.~Gorschek, C.~Wohlin, P.~Carre, and S.~Larsson.
\newblock A {Model} for {Technology} {Transfer} in {Practice}.
\newblock {\em IEEE Software}, 23(6):88--95, 2006.

\bibitem{Grevet2015}
C.~Grevet and E.~Gilbert.
\newblock Piggyback prototyping: Using existing, large-scale social computing
  systems to prototype new ones.
\newblock In {\em Proceedings 33rd Annual ACM Conference on Human Factors in
  Computing Systems}, pages 4047--4056, Seoul, Korea, 2015. ACM.

\bibitem{harb_evaluating_2015}
Y.~Harb, C.~Noteboom, and S.~Sarnikar.
\newblock Evaluating {Project} {Characteristics} for {Selecting} the {Best}-fit
  {Agile} {Software} {Development} {Methodology}: {A} {Teaching} {Case}.
\newblock {\em Journal of the Midwest Association for Information Systems
  (JMWAIS)}, 1(1), 2015.

\bibitem{Hatzakis}
T.~Hatzakis, M.~Lycett, and A.~Serrano.
\newblock {A programme management approach for ensuring curriculum coherence in
  IS (higher) education}.
\newblock {\em European Journal of Information Systems}, 16(5):643--657, 2007.

\bibitem{hayes_research_1999}
W.~Hayes.
\newblock Research synthesis in software engineering: a case for meta-analysis.
\newblock In {\em Proceedings 6th {International} {Software} {Metrics}
  {Symposium}}, pages 143--151, Boca Raton, USA, 1999. IEEE.

\bibitem{Herrmann2015}
B.~L. Herrmann, J.-F. Gauthier, D.~Holtschke, R.~Berman, and M.~Marmer.
\newblock {The Global Startup Ecosystem Ranking 2015}.
\newblock Technical Report August, 2015.

\bibitem{Hill2014}
S.~A. Hill and J.~Birkinshaw.
\newblock {Ambidexterity and Survival in Corporate Venture Units}.
\newblock {\em Journal of Management}, 40(7):1899--1931, 2014.

\bibitem{Hilmola2003}
O.-P. Hilmola, P.~Helo, and L.~Ojala.
\newblock {The value of product development lead time in software startup}.
\newblock {\em System Dynamics Review}, 19(1):75--82, 2003.

\bibitem{hippel_innovation_2001}
E.~v. Hippel.
\newblock Innovation by {User} {Communities}: {Learning} from {Open}-{Source}
  {Software}.
\newblock {\em MIT Sloan Management Review}, 42(4):82--82, 2001.

\bibitem{hokkanen_early_2015}
L.~Hokkanen, K.~Kuusinen, and K.~V{\"a}{\"a}n{\"a}nen.
\newblock Early {Product} {Design} in {Startups}: {Towards} a {UX} {Strategy}.
\newblock In {\em Proceedings 16th {International} {Conference} on
  {Product}-{Focused} {Software} {Process} {Improvement} ({PROFES})}, pages
  217--224, Bolzano-Bozen, Italy, 2015. Springer.

\bibitem{hokkanen_ux_2016}
L.~Hokkanen, K.~Kuusinen, and K.~Väänänen.
\newblock Minimum viable user experience: A framework for supporting product
  design in startups.
\newblock In {\em Proceedings 17th {International} {XP} {Conference} ({XP})},
  Edinburgh, Scotland, 2016. Springer.
\newblock In press.

\bibitem{hokkanen_three_2015}
L.~Hokkanen and M.~Lepp{\"a}nen.
\newblock Three {Patterns} for {User} {Involvement} in {Startups}.
\newblock In {\em Proceedings 20th {European} {Conference} on {Pattern}
  {Languages} of {Programs} ({EuroPLoP})}, pages 51:1--51:8, Kloster Irsee,
  Germany, 2015. ACM.

\bibitem{hokkanen_ux_2015}
L.~Hokkanen and K.~V{\"a}{\"a}n{\"a}nen-Vainio-Mattila.
\newblock {UX} work in startups: current practices and future needs.
\newblock In {\em Proceedings 16th {International} {XP} {Conference} ({XP})},
  pages 81--92, Helsinki, Finland, 2015. Springer.

\bibitem{honig_institutional_2004}
B.~Honig and T.~Karlsson.
\newblock Institutional forces and the written business plan.
\newblock {\em Journal of Management}, 30(1):29--48, 2004.

\bibitem{hu_multi-objective_2008}
G.~Hu, L.~Wang, S.~Fetch, and B.~Bidanda.
\newblock A multi-objective model for project portfolio selection to implement
  lean and {Six} {Sigma} concepts.
\newblock {\em International Journal of Production Research},
  46(23):6611--6625, 2008.

\bibitem{iso_2010}
{International Organization for Standardization}.
\newblock {\em Ergonomics of {Human}-system {Interaction}: {Part} 210:
  {Human}-centred {Design} for {Interactive} {Systems}}.
\newblock ISO, 2010.

\bibitem{jack_small_2006}
S.~Jack, J.~Hyman, and F.~Osborne.
\newblock Small entrepreneurial ventures culture, change and the impact on
  {HRM}: {A} critical review.
\newblock {\em Human Resource Management Review}, 16(4):456--466, 2006.

\bibitem{Jansen2009}
J.~J.~P. Jansen, M.~P. Tempelaar, F.~A.~J. van~den Bosch, and H.~W. Volberda.
\newblock {Structural Differentiation and Ambidexterity: The Mediating Role of
  Integration Mechanisms}.
\newblock {\em Organization Science}, 20(4):797--811, 2009.

\bibitem{Jayshree2012}
S.~Jayshree and R.~Ramraj.
\newblock {Entrepreneurial Ecosystem: Case Study on the Influence of
  Environmental Factors on Entrepreneurial Success}.
\newblock {\em European Journal of Business and Management}, 4(16):95--102,
  2012.

\bibitem{Johne1988114}
F.~Johne and P.~A. Snelson.
\newblock Success factors in product innovation: A selective review of the
  literature.
\newblock {\em Journal of Product Innovation Management}, 5(2):114--128, 1988.

\bibitem{johri_boundary_2008}
A.~Johri.
\newblock Boundary spanning knowledge broker: {An} emerging role in global
  engineering firms.
\newblock In {\em Proceedings 38th {Annual} {Frontiers} in {Education}
  {Conference}}, pages 7--12. IEEE, 2008.

\bibitem{jarvinen_agile_2014}
J.~Järvinen, T.~Huomo, T.~Mikkonen, and P.~Tyrväinen.
\newblock From {Agile} {Software} {Development} to {Mercury} {Business}.
\newblock In {\em Proceedings 5th {International} {Conference} on {Software}
  {Business} ({ICSOB})}, pages 58--71, Paphos, Cyprus, 2014. Springer.

\bibitem{Kamm1990}
J.~Kamm, J.~Shuman, J.~Seeger, and A.~Nurick.
\newblock {Entrepreneurial teams in new venture creation: A research agenda}.
\newblock {\em Entrepreneurship Theory and Practice}, 14(4):7--17, 1990.

\bibitem{karlsson_judging_2009}
T.~Karlsson and B.~Honig.
\newblock Judging a business by its cover: {An} institutional perspective on
  new ventures and the business plan.
\newblock {\em Journal of Business Venturing}, 24(1):27--45, 2009.

\bibitem{katz_properties_1988}
J.~Katz and W.~B. Gartner.
\newblock Properties of {Emerging} {Organizations}.
\newblock {\em Academy of Management Review}, 13(3):429--441, 1988.

\bibitem{Katzenbach1993}
J.~R. Katzenbach and D.~K. Smith.
\newblock {The discipline of teams.}
\newblock {\em Harvard Business Review}, 71(2):111--120, 1993.

\bibitem{kelly_lessons_2012}
M.~D. Kelly.
\newblock Lessons {Learned} from {Software} {Testing} at {Startups}.
\newblock In {\em {EuroStar}-{Software} {Testing} {Conference}}, Amsterdam, The
  Netherlands, 2012.

\bibitem{Kersten2003}
B.~Kersten and C.~Verhoef.
\newblock {IT portfolio management: A banker's perspective on IT}.
\newblock {\em Cutter IT Journal}, 2003.

\bibitem{kirk_categorising_2014}
D.~Kirk and S.~MacDonell.
\newblock Categorising {Software} {Contexts}.
\newblock In {\em Proceedings 2014 {Americas} {Conference} on {Information}
  {Systems} ({AMCIS})}, Savannah, USA, 2014. AIS Electronic Library.

\bibitem{kirk_investigating_2014}
D.~Kirk and S.~G. MacDonell.
\newblock Investigating a {Conceptual} {Construct} for {Software} {Context}.
\newblock In {\em Proceedings 18th {International} {Conference} on {Evaluation}
  and {Assessment} in {Software} {Engineering} ({EASE})}, pages 27:1--27:10,
  London, UK, 2014. ACM.

\bibitem{kirsh_firm_2009}
D.~Kirsh, B.~Goldfarb, and A.~Gera.
\newblock Firm or substance: the role of business plans in venture capital
  decision making process.
\newblock {\em Strategic Management Journal}, (30):487--515, 2009.

\bibitem{kitchenham_robust_2016}
B.~Kitchenham, L.~Madeyski, D.~Budgen, J.~Keung, P.~Brereton, S.~Charters,
  S.~Gibbs, and A.~Pohthong.
\newblock Robust {Statistical} {Methods} for {Empirical} {Software}
  {Engineering}.
\newblock {\em Empirical Software Engineering}, pages 1--52, 2016.

\bibitem{kitchenham_evidence-based_2004}
B.~A. Kitchenham, T.~Dybå, and M.~Jørgensen.
\newblock Evidence-{Based} {Software} {Engineering}.
\newblock In {\em Proceedings 26th {International} {Conference} on {Software}
  {Engineering} ({ICSE})}, pages 273--281, Edinburgh, UK, 2004. IEEE.

\bibitem{klotins_software_2015}
E.~Klotins, M.~Unterkalmsteiner, and T.~Gorschek.
\newblock Software engineering practices in start-up companies: A mapping
  study.
\newblock In {\em 6th International Conference on Software Business}, pages
  245--257. Springer, 2015.

\bibitem{klotins_software_2016}
E.~Klotins, M.~Unterkalmsteiner, and T.~Gorschek.
\newblock Software {Engineering} in {Start}-up {Companies}: an {Exploratory}
  {Study} of 88 {Startups}.
\newblock {\em Empirical Software Engineering}, 2016.
\newblock In Submission.

\bibitem{klyver_resource_2013}
K.~Klyver and M.~T. Schenkel.
\newblock From {Resource} {Access} to {Use}: {Exploring} the {Impact} of
  {Resource} {Combinations} on {Nascent} {Entrepreneurship}.
\newblock {\em Journal of Small Business Management}, 51(4):539--556, 2013.

\bibitem{kon2015}
F.~Kon, D.~Cukier, C.~Melo, O.~Hazzan, and H.~Yuklea.
\newblock A conceptual framework for software startup ecosystems: the case of
  israel.
\newblock Technical report, Technical Report RT-MAC-2015-01, Department of
  Computer Science, University of S\~ao Paulo, 2015.

\bibitem{Krebs2008}
J.~Krebs.
\newblock {\em {Agile portfolio management}}.
\newblock Microsoft Press, 1st edition, 2008.

\bibitem{konig_role_2015}
M.~König, G.~Baltes, and B.~Katzy.
\newblock {On the role of value-network strength as an indicator of
  technology-based venture's survival and growth: Increasing innovation system
  efficiency by leveraging transaction relations to prioritize venture
  support}.
\newblock In {\em Proceedings International Conference on Engineering,
  Technology and Innovation/ International Technology Management Conference
  (ICE/ITMC)}, pages 1--9. IEEE, 2015.

\bibitem{konig_agreement_2016}
M.~König, C.~Ungerer, R.~Büchele, and G.~Baltes.
\newblock Agreement on the {Venture}’s {Reality} {Presented} in {Business}
  {Plans}.
\newblock In {\em Proceedings 22nd {International} {Conference} on
  {Engineering}, {Technology} and {Innovation} ({ICE})}, Trondheim, Norway,
  2016. IEEE.

\bibitem{Lavie2010}
D.~Lavie, U.~Stettner, and M.~L. Tushman.
\newblock {Exploration and Exploitation Within and Across Organizations}.
\newblock {\em The Academy of Management Annals}, 4(1):109--155, 2010.

\bibitem{LeFave2008}
R.~LeFave, B.~Branch, and C.~Brown.
\newblock {How Sprint Nextel Reconfigured IT Resources for Results.}
\newblock {\em MIS Quarterly}, 2008.

\bibitem{Lerner2013corporate}
J.~Lerner.
\newblock Corporate venturing.
\newblock {\em Harvard Business Review}, 91(10):86--94, 2013.

\bibitem{levie_terminal_2010}
J.~Levie and B.~B. Lichtenstein.
\newblock A {Terminal} {Assessment} of {Stages} {Theory}: {Introducing} a
  {Dynamic} {States} {Approach} to {Entrepreneurship}.
\newblock {\em Entrepreneurship Theory and Practice}, 34(2):317--350, 2010.

\bibitem{Li2015}
Z.~Li, P.~Avgeriou, and P.~Liang.
\newblock A systematic mapping study on technical debt and its management.
\newblock {\em Journal of Systems and Software}, 101:193--220, 2015.

\bibitem{lichtenstein_measuring_2006}
B.~B. Lichtenstein, K.~J. Dooley, and G.~T. Lumpkin.
\newblock Measuring emergence in the dynamics of new venture creation.
\newblock {\em Journal of Business Venturing}, 21(2):153--175, 2006.

\bibitem{lichter_prototyping_1993}
H.~Lichter, M.~Schneider-Hufschmidt, and H.~Züllighoven.
\newblock Prototyping in {Industrial} {Software} {Projects}—{Bridging} the
  {Gap} {Between} {Theory} and {Practice}.
\newblock In {\em Proceedings 15th {International} {Conference} on {Software}
  {Engineering} ({ICSE})}, pages 221--229, Baltimore, USA, 1993. IEEE.

\bibitem{Lim2012}
E.~Lim, N.~Taksande, and C.~Seaman.
\newblock A balancing act: What software practitioners have to say about
  technical debt.
\newblock {\em IEEE Software}, 29(6):22--27, 2012.

\bibitem{OECDInnovation}
D.~Lippoldt and P.~Stryszowskim.
\newblock Innovation in the software sector.
\newblock Technical report, Organisation for Economic Co-operation and
  Development, 2009.

\bibitem{lofsten_science_2002}
H.~Löfsten and P.~Lindelöf.
\newblock Science {Parks} and the growth of new technology-based
  firms—academic-industry links, innovation and markets.
\newblock {\em Research Policy}, 31(6):859--876, 2002.

\bibitem{Macmillan1987}
I.~C. Macmillan, L.~Zemann, and P.~Subbanarasimha.
\newblock {Criteria distinguishing successful from unsuccessful ventures in the
  venture screening process}.
\newblock {\em Journal of Business Venturing}, 2(2):123--137, 1987.

\bibitem{marijarvi2016}
J.~M{\"a}rij{\"a}rvi, L.~Hokkanen, M.~Komssi, H.~Kiljander, Y.~Xu,
  M.~Raatikainen, P.~Sepp{\"a}nen, J.~Heininen, M.~Koivulahti-Ojala,
  M.~Helenius, and J.~J{\"a}rvinen.
\newblock {\em The Cookbook for Successful Internal Startups}.
\newblock DIGILE and N4S, 2016.

\bibitem{marlow_human_2006}
S.~Marlow.
\newblock Human resource management in smaller firms: {A} contradiction in
  terms?
\newblock {\em Human Resource Management Review}, 16(4):467--477, 2006.

\bibitem{maurya_running_2012}
A.~Maurya.
\newblock {\em Running {Lean}: {Iterate} from {Plan} {A} to a {Plan} {That}
  {Works}}.
\newblock O'Reilly Media, Inc., 2012.

\bibitem{may_applying_2012}
B.~May.
\newblock Applying {Lean} {Startup}: {An} {Experience} {Report} -- {Lean} \&
  {Lean} {UX} by a {UX} {Veteran}: {Lessons} {Learned} in {Creating} \&
  {Launching} a {Complex} {Consumer} {App}.
\newblock In {\em Proceedings {Agile} {Conference} ({AGILE})}, pages 141--147,
  Dallas, USA, 2012. IEEE.

\bibitem{Meskendahl2010}
S.~Meskendahl.
\newblock {The influence of business strategy on project portfolio management
  and its success - A conceptual framework}.
\newblock {\em International Journal of Project Management}, 28(8):807--817,
  2010.

\bibitem{miles_qualitative_1994}
M.~B. Miles and A.~Huberman.
\newblock {\em Qualitative data analysis: {An} expanded sourcebook}.
\newblock Sage Publications, Inc, Thousand Oaks, US, 1994.

\bibitem{mullins_getting_2009}
J.~W. Mullins and R.~Komisar.
\newblock {\em Getting to {Plan} {B}: {Breaking} {Through} to a {Better}
  {Business} {Model}}.
\newblock Harvard Business Press, 2009.

\bibitem{mulrow_rationale_1994}
C.~D. Mulrow.
\newblock Rationale for systematic reviews.
\newblock {\em BMJ : British Medical Journal}, 309(6954):597--599, 1994.

\bibitem{nambisan_digital_2016}
S.~Nambisan, K.~Lyytinen, A.~Majchrzak, and M.~Song.
\newblock Digital {Innovation} {Management}: {Reinventing} {Innovation}
  {Management} {Research} in a {Digital} {World}.
\newblock {\em MIS Quarterly}, 2016.
\newblock In press.

\bibitem{Nanda2013}
R.~Nanda and M.~Rhodes-Kropf.
\newblock {Investment cycles and startup innovation}.
\newblock {\em Journal of Financial Economics}, 110(2):403--418, 2013.

\bibitem{Nazar2013}
J.~Nazar.
\newblock 14 famous business pivots.
\newblock {\em [AVAILABLE ONLINE]
  http://www.forbes.com/sites/jasonnazar/2013/10/08/14-famous-business-pivots/},
  2013.

\bibitem{newbert_looking_2008}
S.~L. Newbert, S.~Gopalakrishnan, and B.~A. Kirchhoff.
\newblock Looking beyond resources: {Exploring} the importance of
  entrepreneurship to firm-level competitive advantage in technologically
  intensive industries.
\newblock {\em Technovation}, 28(1–2):6--19, 2008.

\bibitem{newbert_supporter_2010}
S.~L. Newbert and E.~T. Tornikoski.
\newblock Supporter networks and network growth: a contingency model of
  organizational emergence.
\newblock {\em Small Business Economics}, 39(1):141--159, 2010.

\bibitem{Newman2015}
P.~Newman, M.~A. Ferrario, W.~Simm, S.~Forshaw, A.~Friday, and J.~Whittle.
\newblock The role of design thinking and physical prototyping in social
  software engineering.
\newblock {\em 37th IEEE International Conference on Software Engineering},
  2015.

\bibitem{NguyenDuc2016}
A.~Nguyen~Duc and P.~Abrahamsson.
\newblock Minimum viable product or multiple facet product? {T}he {R}ole of
  {MVP} in software startups.
\newblock In {\em Proceedings 17th International XP Conference}, Edinburgh, UK,
  2016. Springer.

\bibitem{NguyenDuc2015}
A.~Nguyen~Duc, P.~Sepp{\"a}nen, and P.~K. Abrahamsson.
\newblock Hunter-gatherer cycle: a conceptual model of the evolution of
  software startups.
\newblock pages 199--203, Tallin, Estonia, 2015. ACM.

\bibitem{nobel_teaching_2011}
C.~Nobel.
\newblock Teaching a ‘{Lean} {Startup}’ {Strategy}.
\newblock {\em HBS Working Knowledge}, 2011.

\bibitem{Oechslein2012}
O.~Oechslein and A.~Tumasjan.
\newblock {Examining Trust within the Team in IT Startup Companies--An
  Empirical Study in the People's Republic of China}.
\newblock In {\em Proceedings 45th Hawaii International Conference on System
  Science (HICSS)}, pages 5102--5111, Maui, USA, 2012.

\bibitem{OReilly2013}
C.~A. O'Reilly and M.~L. Tushman.
\newblock {Organizational Ambidexterity: Past, Present, and Future}.
\newblock {\em Academy of Management Perspectives}, 27(4):324--338, 2013.

\bibitem{paternoster_software_2014}
N.~Paternoster, C.~Giardino, M.~Unterkalmsteiner, T.~Gorschek, and
  P.~Abrahamsson.
\newblock Software {Development} in {Startup} {Companies}: {A} {Systematic}
  {Mapping} {Study}.
\newblock {\em Information and Software Technology}, 56(10):1200--1218, 2014.

\bibitem{pelrine_understanding_2011}
J.~Pelrine.
\newblock On {Understanding} {Software} {Agility}: {A} {Social} {Complexity}
  {Point} {Of} {View}.
\newblock {\em Emergence: Complexity \& Organization}, 13(1/2):26--37, 2011.

\bibitem{petersen_context_2009}
K.~Petersen and C.~Wohlin.
\newblock Context in {Industrial} {Software} {Engineering} {Research}.
\newblock In {\em Proceedings 3rd {International} {Symposium} on {Empirical}
  {Software} {Engineering} and {Measurement} ({ESEM})}, pages 401--404,
  Orlando, USA, 2009. IEEE.

\bibitem{pikkarainen_art_2011}
M.~Pikkarainen, W.~Codenie, N.~Boucart, and J.~A. Heredia~Alvaro, editors.
\newblock {\em The {Art} of {Software} {Innovation}}.
\newblock Springer, Berlin, Germany, 2011.

\bibitem{raz_use_2001}
T.~Raz and E.~Michael.
\newblock Use and benefits of tools for project risk management.
\newblock {\em International Journal of Project Management}, 19(1):9--17, 2001.

\bibitem{Reyck2005}
B.~D. Reyck, Y.~Grushka-Cockayne, M.~Lockett, S.~R. Calderini, M.~Moura, and
  A.~Sloper.
\newblock {The impact of project portfolio management on information technology
  projects}.
\newblock {\em International Journal of Project Management}, 23(7):524--537,
  2005.

\bibitem{ries_lean_2011}
E.~Ries.
\newblock {\em The lean startup: {How} today's entrepreneurs use continuous
  innovation to create radically successful businesses}.
\newblock Crown Books, 2011.

\bibitem{rikkila_implications_2012}
J.~Rikkila, P.~Abrahamsson, and X.~Wang.
\newblock The {Implications} of a {Complexity} {Perspective} for {Software}
  {Engineering} {Practice} and {Research}.
\newblock {\em Journal of Computer Engineering \& Information Technology},
  2012.

\bibitem{Sarasvathy2001}
S.~D. Sarasvathy.
\newblock {Causation and Effectuation: Toward a Theoretical Shift from Economic
  Inevitability to Entrepreneurial Contingency}.
\newblock {\em Academy of Management}, 26(2):243--263, 2001.

\bibitem{semrau_networking_2012}
T.~Semrau and S.~Sigmund.
\newblock Networking {Ability} and the {Financial} {Performance} of {New}
  {Ventures}: {A} {Mediation} {Analysis} among {Younger} and {More} {Mature}
  {Firms}.
\newblock {\em Strategic Entrepreneurship Journal}, 6(4):335--354, 2012.

\bibitem{ShahidBajwa2016}
S.~Shahid~Bajwa, X.~Wang, A.~Nguven~Duc, and P.~Abrahamsson.
\newblock {How Do Software Startups Pivot? Empirical Results from a Multiple
  Case Study}.
\newblock In {\em 7th International Conference on Software Business (ICSOB
  2016)}, pages 169--176, Ljubljana, Slovenia, 2016.

\bibitem{Shontell2012}
A.~Shontell.
\newblock The 11 most disruptive startups.
\newblock {\em Business Insider}, 07/12 2012.

\bibitem{shull_technical_2013}
F.~Shull, D.~Falessi, C.~Seaman, M.~Diep, and L.~Layman.
\newblock Technical {Debt}: {Showing} the {Way} for {Better} {Transfer} of
  {Empirical} {Results}.
\newblock In {\em Perspectives on the {Future} of {Software} {Engineering}},
  pages 179--190. Springer, 2013.

\bibitem{snowden_leaders_2007}
D.~J. Snowden and M.~E. Boone.
\newblock A leader's framework for decision making.
\newblock {\em Harvard Business Review}, 85(11):69--76, 2007.

\bibitem{sommerville_software_2010}
I.~Sommerville.
\newblock {\em Software {Engineering}}.
\newblock Pearson, Boston, 9th edition, 2010.

\bibitem{souza_guest_2009}
C.~R. B.~d. Souza, H.~Sharp, J.~Singer, L.~T. Cheng, and G.~Venolia.
\newblock Guest {Editors}' {Introduction}: {Cooperative} and {Human} {Aspects}
  of {Software} {Engineering}.
\newblock {\em IEEE Software}, 26(6):17--19, 2009.

\bibitem{srinivasan_venture_2014}
S.~Srinivasan, I.~Barchas, M.~Gorenberg, and E.~Simoudis.
\newblock Venture {Capital}: {Fueling} the {Innovation} {Economy}.
\newblock {\em Computer}, 47(8):40--47, 2014.

\bibitem{steinert_finding_2012}
M.~Steinert and L.~J. Leifer.
\newblock ‘{Finding} {One}’s {Way}’: {Re}-{Discovering} a {Hunter} -
  {Gatherer} {Model} based on {Wayfaring}.
\newblock {\em International Journal of Engineering Education}, 28(2):251--252,
  2012.

\bibitem{Sternberg2013}
R.~Sternberg.
\newblock {Success factors of university-spin-offs: Regional government support
  programs versus regional environment}.
\newblock {\em Technovation}, pages 1--12, 2013.

\bibitem{sutton_role_2000}
S.~M. Sutton.
\newblock The {Role} of {Process} in a {Software} {Start}-up.
\newblock {\em IEEE Softw.}, 17(4):33--39, 2000.

\bibitem{taipale_huitalestory_2010}
M.~Taipale.
\newblock Huitale--{A} {Story} of a {Finnish} {Lean} {Startup}.
\newblock In {\em Proceedings 1st {International} {Conference} on {Lean}
  {Enterprise} and {Software} {Systems} ({LESS})}, pages 111--114, Helsinki,
  Finland, 2010. Springer.

\bibitem{teece_dynamic_1997}
D.~J. Teece, G.~Pisano, and A.~Shuen.
\newblock Dynamic capabilities and strategic management.
\newblock {\em Strategic Management Journal}, 18(7):509--533, 1997.

\bibitem{Terho2015}
H.~Terho, S.~Suonsyrj, A.~Karisalo, and T.~O. Mikkonen.
\newblock {Ways to cross the rubicon: Pivoting in software startups}.
\newblock In {\em Proceedings 1st International Workshop on Software Startups},
  pages 555--568, Bolzano-Bozen, Italy, 2015. Springer.

\bibitem{Economist20140118}
{The Economist}.
\newblock A cambrian moment cheap and ubiquitous building blocks for digital
  products and services have caused an explosion in startups. special report:
  Tech startups.
\newblock {\em The Economist}, 01/18 2014.

\bibitem{Economist20140118-2}
{The Economist}.
\newblock Testing, testing launching a startup has become fairly easy, but what
  follows is back-breaking work. special report: Tech startups.
\newblock {\em The Economist}, 01/18 2014.

\bibitem{Economist20150919}
{The Economist}.
\newblock Progress without profits. a flock of startups is making cloud
  computing faster and more flexible, but most of them will not survive.
\newblock {\em The Economist}, 09/19 2015.

\bibitem{tom_exploration_2013}
E.~Tom, A.~Aurum, and R.~Vidgen.
\newblock An exploration of technical debt.
\newblock {\em Journal of Systems and Software}, 86(6):1498--1516, 2013.

\bibitem{torkar_adopting_2011}
R.~Torkar, P.~Minoves, and J.~Garrigós.
\newblock Adopting free/libre/open source software practices, techniques and
  methods for industrial use.
\newblock {\em Journal of the Association for Information Systems}, 12(1):88,
  2011.

\bibitem{turner2014handbook}
J.~R. Turner.
\newblock {\em The handbook of project-based management}.
\newblock McGraw-hill, 2014.

\bibitem{tyrvainen_metrics_2015}
P.~Tyrväinen, M.~Saarikallio, T.~Aho, T.~Lehtonen, and R.~Paukeri.
\newblock Metrics {Framework} for {Cycle}-{Time} {Reduction} in {Software}
  {Value} {Creation}.
\newblock In {\em Proceedings 10th {International} {Conference} on {Software}
  {Engineering} {Advances} ({ICSEA})}, Barcelona, Spain, 2015. IARIA.

\bibitem{ungerer_measuring_2016}
C.~Ungerer, M.~König, F.~Giones, and G.~Baltes.
\newblock Measuring {Venture} {Emergence} and {Survival} by {Analyzing}
  {Transaction} {Relations} in {Business} {Plans}.
\newblock In {\em Proceedings 22nd {International} {Conference} on
  {Engineering}, {Technology} and {Innovation} ({ICE})}, Trondheim, Norway,
  2016. IEEE.

\bibitem{unterkalmsteiner_conceptual_2014}
M.~Unterkalmsteiner, T.~Gorschek, A.~Islam, C.~Cheng, R.~Permadi, and R.~Feldt.
\newblock A conceptual framework for {SPI} evaluation.
\newblock {\em Journal of Software: Evolution and Process}, 26(2):251--279,
  2014.

\bibitem{VandeVrande2009}
V.~van~de Vrande, J.~P. de~Jong, W.~Vanhaverbeke, and M.~de~Rochemont.
\newblock {Open innovation in SMEs: Trends, motives and management challenges}.
\newblock {\em Technovation}, 29(6):423--437, 2009.

\bibitem{VanderVen2013}
J.~S. van~der Ven and J.~Bosch.
\newblock {Pivots and Architectural Decisions: Two Sides of the Same Medal?}
\newblock In {\em Proceedings 8th International Conference on Software
  Engineering Advances (ICSEA)}, pages 310--317, Venice, Italy, 2013.

\bibitem{Vanhaverbeke2006}
W.~Vanhaverbeke and M.~Cloodt.
\newblock {Open innovation in value networks}.
\newblock {\em Open innovation: Researching a New Paradigm}, 2006.

\bibitem{ven_explaining_1995}
A.~H. V.~D. Ven and M.~S. Poole.
\newblock Explaining {Development} and {Change} in {Organizations}.
\newblock {\em Academy of Management Review}, 20(3):510--540, 1995.

\bibitem{wenger_communities_1998}
E.~Wenger.
\newblock {\em Communities of {Practice}: {Learning}, {Meaning}, and
  {Identity}}.
\newblock Cambridge University Press, 1998.

\bibitem{west_how_2003}
J.~West.
\newblock How open is open enough?: {Melding} proprietary and open source
  platform strategies.
\newblock {\em Research Policy}, 32(7):1259--1285, 2003.

\bibitem{West2006}
J.~West and S.~Gallagher.
\newblock {Challenges of open innovation: the paradox of firm investment in
  open-source software}.
\newblock {\em R \& D Management}, 36(3):319--331, 2006.

\bibitem{wieringa_empirical_2014}
R.~Wieringa.
\newblock Empirical research methods for technology validation: {Scaling} up to
  practice.
\newblock {\em Journal of Systems and Software}, 95:19--31, 2014.

\bibitem{williams_competent_2002}
P.~Williams.
\newblock The {Competent} {Boundary} {Spanner}.
\newblock {\em Public Administration}, 80(1):103--124, 2002.

\bibitem{witt_entrepreneurs_2004}
P.~Witt.
\newblock Entrepreneurs’ networks and the success of start-ups.
\newblock {\em Entrepreneurship \& Regional Development}, 16(5):391--412, 2004.

\bibitem{wmf2015}
WMF.
\newblock Intelligent assets unlocking the circular economy potential.
\newblock Technical report, World Economic Forum, December 2015.

\bibitem{WohlinClaes2012}
J.~{Wohlin, Claes; Aurum, Aybuke; Angelis, Lefteris; Phillips, Laura; Dittrich,
  Yvonne; Gorschek, Tony; Grahn, Hakan; Henningsson, Kennet; Kagstrom, Simon;
  Low, Graham; Rovegard, Per; Tomaszewski, Piotr; van Toorn, Christine;
  Winter}.
\newblock {The Success Factors Powering Academia Collaboration}.
\newblock {\em IEEE Software}, 29(2):67--73, 2012.

\bibitem{yague_analyzing_2014}
A.~Yag{\"u}e, J.~Garbajosa, J.~P{\'e}rez, and J.~D{\'\i}az.
\newblock Analyzing {Software} {Product} {Innovation} {Assessment} by {Using} a
  {Systematic} {Literature} {Review}.
\newblock In {\em Proceedings 47th {Hawaii} {International} {Conference} on
  {System} {Sciences} ({HICSS})}, pages 5049--5058, Waikoloa, USA, 2014. IEEE.

\bibitem{yau_is_2013}
A.~Yau and C.~Murphy.
\newblock Is a {Rigorous} {Agile} {Methodology} the {Best} {Development}
  {Strategy} for {Small} {Scale} {Tech} {Startups}?
\newblock Technical {Report} MS-CIS-13-01, 2013.

\bibitem{yin_case_2003}
R.~K. Yin.
\newblock {\em Case {Study} {Research}: {Design} and {Methods}}.
\newblock Sage Publications, 3rd edition, 2003.

\bibitem{zettell_lipe:_2001}
J.~Zettel, F.~Maurer, J.~Münch, and L.~Wong.
\newblock {LIPE}: a lightweight process for e-business startup companies based
  on extreme programming.
\newblock In {\em Proceedings 3rd International Conference on Product-Focused
  Software Process Improvement (PROFES)}, pages 255--270. Springer,
  Kaiserslautern, Germany, 2001.

\end{thebibliography}
\bibliographystyle{abbrv}
	
\end{document}